\documentclass[10pt,aps,pra,twocolumn,superscriptaddress,longbibliography,nobalancelastpage,floatfix]{revtex4-2}
\usepackage[T1]{fontenc}
\usepackage[utf8]{inputenc}
\pdfoutput=1 
\usepackage{graphicx}
\usepackage{mathtools}
\usepackage{bm}
\usepackage{physics}
\usepackage[colorlinks = true,
            allcolors = {Blue}]{hyperref}
\usepackage[capitalize]{cleveref}
\usepackage[caption=false]{subfig}
\newcommand{\phantomsubfloat}[1]{
    {%
        \captionsetup[subfigure]{labelformat=empty}
        \subfloat[][]{#1}
    }%
}
\usepackage[dvipsnames]{xcolor}

\usepackage[normalem]{ulem}

\newcommand{\pp}{{p^{\prime}}}
\newcommand{\x}{{\rm x}}
\newcommand{\y}{{\rm y}}
\makeatletter
\newcommand{\supplementarytocfile}{stoc}

\let\oldsection\section
\renewcommand{\section}{%
  \@ifstar{\oldsection*}{\suppsection}%
}

\newcommand{\suppsection}[1]{%
  \oldsection{#1}%
  \addcontentsline{\supplementarytocfile}{section}{\protect\numberline{\thesection}#1}%
}

\def\l@section#1#2{\addpenalty\@secpenalty\addvspace{0pt}%
  \setlength\@tempdima{2.5em}%
  \begingroup
    \parindent \z@
    \rightskip \@pnumwidth
    \parfillskip -\@pnumwidth
    \leavevmode
    \advance\leftskip\@tempdima
    \hskip -\leftskip
    #1\nobreak\hfil\nobreak\hb@xt@\@pnumwidth{\hss #2}\par
  \endgroup
}
\makeatother

\begin{document}

\title{Mapping molecular polariton transport via pump-probe microscopy}

\author{Piper Fowler-Wright}
\affiliation{Department of Chemistry and Biochemistry, University of California San Diego, La Jolla, California 92093, USA}
\author{Michael Reitz}
\affiliation{Department of Chemistry and Biochemistry, University of California San Diego, La Jolla, California 92093, USA}
\author{Joel Yuen-Zhou}
\affiliation{Department of Chemistry and Biochemistry, University of California San Diego, La Jolla, California 92093, USA}

\date{\today}

\begin{abstract}

We demonstrate how the transport properties of molecular polaritons in optical
cavities can be extracted from a microscopic modeling of pump-probe
spectroscopy. Our approach combines a mean-field treatment of the light-matter
Hamiltonian with a perturbative expansion of both light and matter components,
along with spatial coarse-graining. This approach extends semiclassical cavity
spectroscopy to multimode light--matter interactions, providing full access to
spatially resolved transient spectra. By simulating a microscopy experiment with
counter-propagating pump and probe pulses, we compute the differential
transmission and show how molecular dephasing and persistent dark exciton
populations drive sub-group-velocity transport of the root-mean-square
displacement. We analyze transport across the polariton dispersion, showing how
velocity renormalization correlates with excitonic weight, consistent with
experimental observations, and further its dependence on the rate of molecular
dephasing, exciton hopping, and exciton-exciton annihilation. Our results
highlight the need to consider measured spectroscopic observables when
characterizing transport in polaritonic systems.

\vspace{0.5em}
\end{abstract}

\maketitle

Exciton polaritons are hybrid light-matter quasiparticles formed through strong
coupling of confined electromagnetic modes with electronic excitations. Their
photonic component allows them to travel ballistically with a small effective
mass, enabling efficient, long-range energy transport that may be exploited in
novel optoelectronic devices~\cite{lerario2016,hou2020,liu2023}.  Molecular
materials such as organic dye films offer a particularly promising platform for
this purpose, on account of the ease and flexibility of their fabrication, and
operability at room temperatures~\cite{ghosh2022}.

Recent advancements in ultrafast nonlinear spectroscopic techniques have enabled
imaging of molecular polariton transport on sub-picosecond timescales in a range
of closed and open cavity
structures~\cite{rozenman2018,balasubrahmaniyam2023,berghuis2022,berghuis2024,
pandya2021,pandya2022,rozenman2018,xu2023,jin2023, cheng2023}.  Surprisingly,
these experiments report not simply ballistic transport at the polariton group
velocity, but transport velocities renormalized according to the excitonic
content of the propagating excitation, and, ultimately, a crossover to diffusive
behavior.  These results have prompted extensive
research~\cite{groenhof2019,suyabatmaz2023,osipov2023,sokolovskii2023,
tichauer2023,tutunnikov2024,zhou2024,ying2024,krupp2024,liu2024,blackham2025,chng2025}
into the mechanism of transport of hybrid light-matter states in molecular
systems  and, in particular, the role of dark exciton states in the
dynamics~\cite{khazanov2023}.

A large number of approaches have been taken to model molecular polariton
transport, ranging from kinetic models of thermally-activated
scattering~\cite{balasubrahmaniyam2023} to mixed quantum-classical
treatments~\cite{xu2023,koshkaki2025} and molecular dynamics
simulations~\cite{groenhof2019,sokolovskii2023,tichauer2023}.  While these
models capture different aspects of the transport, they do so by characterizing
individual photon, molecular, or polariton populations---quantities that may not
directly correspond to experimental observables: in the time-resolved technique
of pump-probe microscopy, transient signals are often collected without regard
as to their specific origin, polaritonic or
otherwise~\cite{khazanov2023,renken2021}. A more complete interpretation thus
requires understanding how different excitations contribute to the observed
spectroscopic signal.

\begin{figure}[t]
    \centering
    \includegraphics[width=\linewidth]{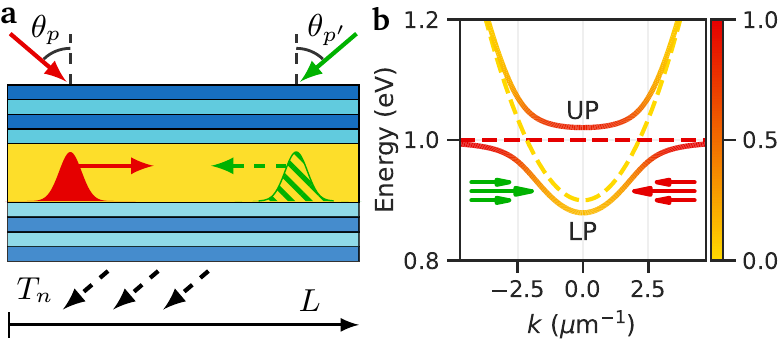}
    \phantomsubfloat{\label{fig:1a}}
    \phantomsubfloat{\label{fig:1b}}
	\vspace{-\baselineskip}%
    \caption{\textit{Schematics.} 
	(a) A planar microcavity of extent $L$ is driven by counter-propagating pump
	(\(p\)) and probe (\(\pp\)) fields impinging at \(\theta_{p,\pp}\).  The
	pump generates laterally-propagating light-matter wavepackets probed by the
	\(\pp\) field.  Information of pump-induced transport is retrieved through
	the spatially-resolved transmitted field \(T_n\) detected along the probe
	direction. 
	(b) Polariton dispersions versus in-plane momentum \(k\). The color scale
	indicates the excitonic fraction \(X_k^2\) of the upper (UP) and lower (LP)
	polariton branches and the  bare cavity dispersion  and exciton energy are
	shown as dashed lines.  Arrows indicate resonant excitation of LP by the
	pump and probe pulses centered at \(k_{p,\pp}=\pm \pi/2\ \mu\text{m}^{-1}\).
    }
    \label{fig:1}
\end{figure}

In this Letter, we present a microscopic model of a pump-probe transport
experiment in an organic microcavity (\cref{fig:1a}).  Our method extends the
framework for nonlinear cavity spectroscopy introduced in
Ref.~\cite{reitz2025nonlinear}, which is based on a perturbative expansion of
both light and matter components of the system, to multimode configurations.
This allows us to gain insight into the different contributions to spectroscopic
signals in a space- and time-resolved manner.  We apply this method to show how
sub-group-velocity transport, as extracted from the root-mean-square
displacement, emerges in a simple model of molecular polariton transport with
pure dephasing.  In particular, we assess the role of dark excitonic populations
in renormalizing the velocity and the dependence of transport properties on
in-plane cavity momentum and dephasing rate.\\

We consider \(N\)  molecules as a
collection of two-level systems (Pauli matrices \(\sigma^\alpha_i\)) interacting
with \(N_k\) modes of a microcavity (bosonic annihilation operators
\(a^{\vphantom{\dagger}}_{k}\)) according to the Tavis--Cummings (TC)
Hamiltonian under the rotating-wave approximation.  
Setting~\footnote{See Supporting Information Table 1 for relation to physical units
used for results presented in the letter.} \(\hbar=1\),
\begin{align}
\begin{split}
	H_{} &= \sum_{k=1}^{N_k}
	{\omega}^{\vphantom{\dagger}}_{k} a^{\dagger}_{k}a^{\vphantom{\dagger}}_{k}
    +  \sum_{n=1}^{N} \frac{\omega_0^{\vphantom{\dagger}}}{2} \sigma^z_{n} \\
	&+   \sum_{k=1}^{N_k}\sum_{n=1}^{N}
	g(a^{\vphantom{\dagger}}_{k}\sigma^+_{n}e^{ikr_n} + 
    \text{H.c.}),
\end{split}
\label{eq:H}
\end{align}
where \({\omega}^{\vphantom{\dagger}}_{0}\) is the molecular transition
frequency,  \(g\) the light-matter coupling strength of a single molecule, and
\({\omega}^{\vphantom{\dagger}}_{k}\)  describes the cavity dispersion. 

For simplicity, we consider a 1D  chain where the \(N\) molecules
are distributed between \(N_k\) sites, and a quadratic dispersion
${\omega}^{\vphantom{\dagger}}_{k} =  {\omega}^{\vphantom{\dagger}}_{c}  + k^2
c^2/2 {\omega}^{\vphantom{\dagger}}_{c}$,   which approximates the dispersion of
a planar microcavity  ~\cite{tichauer2021}. Here, \(k\) is the in-plane
momentum,  quantized as \(k = 2\pi Q / L\) with  \(Q = -N_k/2, \dots, N_k/2\)
and system length \(L\); \(N_k\) is chosen such that \(\Delta r = L/N_k\)
provides adequate spatial resolution.  The cavity frequency
\({\omega}^{\vphantom{\dagger}}_{c}\) is set by the normal wavevector
component, \(k_\perp = {\omega}^{\vphantom{\dagger}}_{c} / c\), in turn fixed by
the resonance condition of the relevant longitudinal cavity mode.  

Note that we have placed no particular restriction on the type of molecular matter in the model; \cref{eq:H} describes 
with generality the coupling between quantum emitters and collective cavity modes,
as relevant to a wide
range of organic microcavity systems including, for example, molecular 
aggregates~\cite{liu2020}, semiconductors~\cite{balasubrahmaniyam2023} 
and halide perovskites~\cite{xu2023}. Whilst the Tavis-Cummings Hamiltonian
may also be applied to inorganic systems across cavity-QED, here our
focus is on molecular materials where vibronic coupling gives rise to 
dephasing and dynamical disorder---processes central to the 
transport mechanisms we investigate. Further, while we assume identical 
molecules, i.e., uniform \({\omega}^{\vphantom{\dagger}}_{0}\)
and \(g\), the calculations presented below can be readily extended to include energetic disorder and inhomogeneous couplings, as shown in the Supporting Information (SI).

Explicit expressions for the eigenstates of \cref{eq:H} may be derived for low
excitation densities~\cite{fowler-wright2024a}.  At each \(k\), the system
supports two bright eigenstates, the upper (UP, \(+\)) and lower (LP, \(-\))
polaritons, with frequencies
\begin{align}
    \omega^{\text{UP/LP}}_k =  
    \frac{1}{2}\left[ 
            {\omega}^{\vphantom{\dagger}}_{0} + {\omega}^{\vphantom{\dagger}}_{k}
            \pm  \sqrt{({\omega}^{\vphantom{\dagger}}_{0}
	- {\omega}^{\vphantom{\dagger}}_{k})^2+ 4 \Omega^2}\right].
    \label{eq:epLepU}
\end{align}
Here \(\Omega = g\sqrt{N}\)  represents the collective light-matter coupling,
yielding a splitting  of $2\Omega$ at resonance.

The frequencies \(\omega^{\text{UP/LP}}_k\) are plotted alongside the  cavity
dispersion and exciton frequency in \cref{fig:1b}.  Importantly, the group
velocity  \(v^{\text{grp}}_k = \partial_k^{\phantom{\dagger}}
\omega^{\text{LP}}_k\) provides a theoretical reference for  comparing with the
observed velocities in our simulations following LP excitation.  In addition,
the excitonic fraction \(X_k^2\) of the  eigenstates, shown as a color scale in
\cref{fig:1b},  will be useful to connect excitation composition to transport
properties.  Broadly, as \(X_k^2\) increases with \(k > 0\) along  the lower
polariton branch, one expects more matter-like, i.e., slowly diffusive,
behavior.

After the polaritons, the remaining degrees of freedom comprise asymmetric
molecular  combinations at the transition frequency \(
{\omega}^{\vphantom{\dagger}}_{0}  \) that are optically
dark~\cite{khazanov2023}. As we explain below, the population of these dark
exciton states, which are stationary (or slowly diffusing), is critical to the
observed transport.

In addition to the unitary dynamics generated by \cref{eq:H},
we consider decay at rate \(\kappa\) from each cavity mode and dephasing 
\(\gamma_\phi\) of the molecules. These are included as Markovian loss terms
in the master equation for the total density operator \(\rho\),
\begin{align}
	\begin{split}
		\partial_t \rho = &-i \bigl[ H , \rho \bigr]
	+ \sum_{k=1}^{N_k} \kappa \mathcal{L}[a^{\vphantom{\dagger}}_{k}]
    + \sum_{n=1}^{N} 
    \frac{\gamma_\phi}{4}\mathcal{L}[\sigma^z_n],
\end{split}
	\label{eq:ME}
\end{align}
with \(\mathcal{L}[x]=x\rho x^\dagger - \{x^\dagger x, \rho\}/2\). 

Eq.~\eqref{eq:ME} provides a minimal model to demonstrate the essential features
of polariton transport under simple dephasing processes. In the SI,
we include an extended model including
exciton hopping and exciton-exciton annihilation processes.  More complex
descriptions of molecular and photonic environments may also be incorporated.\\

 To solve the model described
by \cref{eq:ME}, we apply a mean-field  approximation, which assumes that the
density operator \(\rho\) has a product form.  Expectation values involving
photonic and molecular operators can then be factorized, e.g.,  \(\langle
a^{\vphantom{\dagger}}_{k} \sigma_n^z \rangle \approx  \langle
a^{\vphantom{\dagger}}_{k} \rangle \langle \sigma_n^z \rangle\).  This ansatz is
exact for the TC model in the limit  \(N \to \infty\)~\cite{carollo2021} and
produces a closed set of Heisenberg  equations of motion for the operators
\(\langle a^{\phantom{\dagger}}_{k}\rangle\),  \(\langle \sigma^-_n\rangle\) and
\(\langle\sigma^z_n\rangle\).  The number of equations scales with the number of
chain positions \(N_k\), with \(N_k\sim10^2\) for micrometer resolution for the
system length \(L=200\, \mu \text{m}\) we consider. Deviation from mean-field
behavior may occur when cavity vacuum fluctuations become relevant, e.g., in
spontaneous photoluminescence measurements \cite{chen2019}. However, we do not
expect these processes to be important for the nonlinear microscopy signals of
interest.

Introducing the transform \( a^{\vphantom{\dagger}}_{n}=(1/\sqrt{N_k}) \sum_{k}
e^{ikr_n}a^{\vphantom{\dagger}}_{k}\) of the photon operators, the mean-field
equations can be cast in real space as
\begin{subequations}
\label{eq:mf}
\begin{align}
\begin{split}
    		\partial_t \langle \tilde{a}^{\vphantom{\dagger}}_{n}\rangle  &=
- \bigl(i {\omega}^{\vphantom{\dagger}}_{c}+ \kappa/2\bigr)
\langle \tilde{a}^{\vphantom{\dagger}}_n \rangle  
- i \Omega  \langle \sigma^-_{n} \rangle  \\
&\phantom{=}\
+iC \partial_n^2 \langle \tilde{a}^{\vphantom{\dagger}}_{n} \rangle
+ \sqrt{\kappa} \langle \tilde{a}_n^{\text{in}} \rangle,
\end{split}\label{eq:mfa}\\
		\partial_t \langle \sigma^-_{n} \rangle  &= -
		\bigl( i \omega_0^{\vphantom{\dagger}} +
\gamma_\phi/2 \bigr) \langle \sigma^-_{n} \rangle
+ i \Omega
\langle \sigma^z_{n} \rangle \langle \tilde{a}^{\vphantom{\dagger}}_{n} \rangle ,
		\label{eq:mfb}	
		\\
		\partial_t \langle \sigma^z_{n} \rangle &= 
		- 4 \Omega
		\Im \left[ \langle \sigma^-_{n} \rangle\langle
				\tilde{a}^{\vphantom{\dagger}}_{n}\rangle^*
		\right],\label{eq:mfc}
\end{align}
\end{subequations}
where \(C=c^2/(2{\omega}^{\vphantom{\dagger}}_{c}\Delta r^2)\), and we defined
the rescaled photon variables \(\tilde{a}^{\vphantom{\dagger}}_{n}
=\sqrt{N_k/N}a^{\vphantom{\dagger}}_{n}\), such that the equations do not depend
explicitly on \(N\). A derivation of \crefrange{eq:mfa}{eq:mfc} from
\cref{eq:ME} is provided in the SI. The input field \(\langle
\tilde{a}_n^{\text{in}} \rangle\) is defined further below. The mean-field
equations can be extended in a straightforward fashion to include processes that
may become important in real materials such as exciton hopping and
exciton-exciton annihilation (see SI).

The mean-field approach with coarse-graining provides an efficient way of
computing the spatially-resolved dynamics of microcavity systems with large
numbers of molecules, without  restriction to the first-excitation manifold.
Next, we explain how the method can be  integrated with the perturbative
nonlinear spectroscopy framework of  Ref.~\cite{reitz2025nonlinear} to compute
spatially-resolved pump-probe spectra.

In pump-probe microscopy experiments, two excitation  schemes are commonly
used~\cite{xu2023,khazanov2023}.  In the first, non-resonant pump at high
energies creates a reservoir of  electronic excitations that scatter and
populate the polariton  branches indiscriminately to a large degree. A probe
pulse is then tuned to  a target energy and momentum. In the second, the pump is
itself tuned  to a specific point on the polariton dispersion,  generating
coherent polariton populations.  This second, resonant excitation scheme, can be
more naturally incorporated in the mean-field approach.

Pump and probe pulses are included as driving terms for the photonic variable
\(\langle \tilde{a}^{\vphantom{\dagger}}_{n} \rangle\) in \cref{eq:mfa}, by
setting
\begin{align}
    \sqrt{\kappa} \langle \tilde{a}^{\text{in}}_n (t) \rangle=
    \eta_{p} f_{p}(t)  
      e^{-i  {\omega}^{\vphantom{\dagger}}_{p} t}D_n^{p} 
      +
     \eta_{p'}  f_{p'}(t)  
      e^{-i  {\omega}^{\vphantom{\dagger}}_{p'} t} 
      D_n^{p'}.\label{eq:ain}
\end{align}
Here, \(\eta_{\beta}\), \(f_{\beta}(t)\) and
\({\omega}^{\vphantom{\dagger}}_{{\beta}}\) define  the amplitude, temporal
pulse shape and frequency  of the pump (\(\beta=p\)) and probe
(\(\beta=p'\)), while \(D_n^{\beta}\) specifies their spatial profiles.
The profiles include a modulation \(\sim e^{ik_{\beta} r_n}\) at the central
pump and probe wavevectors, which we initially set to \(k_{p}=-k_{p'} =
\pi/2\)~\(\mu\text{m}^{-1}\). 
Both pump and probe are therefore introduced on equal footing at the level of the driving terms, while their distinct physical roles are resolved by a perturbative expansion in amplitudes 
made below (cf.~\cref{eq:expansion}).

Below, we consider Gaussian pulses of temporal width  \(\sigma_t = 25\)~fs and
spatial width \(\sigma_n = 5\)~\(\mu\)m centered in energy on the lower
polariton frequency at the target momenta,
\({\omega}^{\vphantom{\dagger}}_{p}={\omega}^{\vphantom{\dagger}}_{p'}=\omega^{\text{LP}}_{k_p}\).
Corresponding expressions for \(f_\beta(t)\) and \(D_n^\beta\) are given in the
SI.\\

 Before analyzing the
spectroscopic signal observed in pump-probe experiments, we examine pump-only
dynamics, i.e., \(\eta_{p'}=0\), which provide insight into  the role of dark
populations in transport.

\begin{figure}
    \centering
    \includegraphics[width=\linewidth]{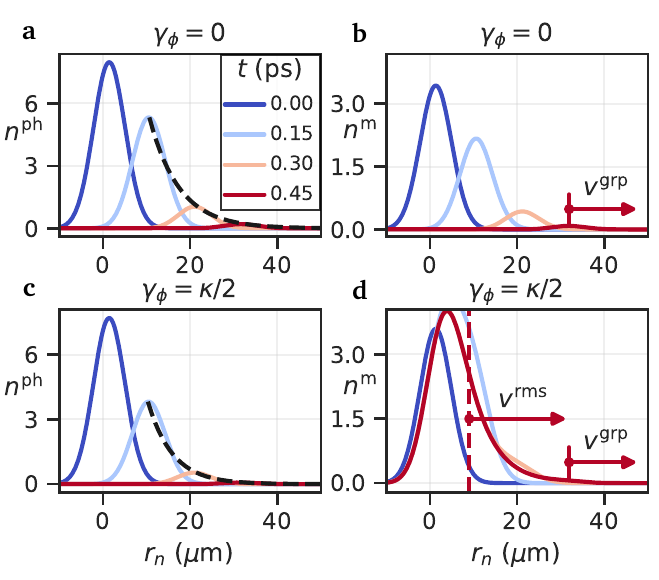}
    \phantomsubfloat{\label{fig:2a}}%
    \phantomsubfloat{\label{fig:2b}}%
    \phantomsubfloat{\label{fig:2c}}%
    \phantomsubfloat{\label{fig:2d}}%
	\vspace{-2\baselineskip}%
    \caption{
    \textit{Pump-induced population dynamics.}   
 Dynamics of  total (a) photon and (b) molecular populations in the absence of
 dephasing (\(\gamma_\phi=0\)).  (c), (d) Corresponding dynamics when dephasing
 is introduced (\(\gamma_\phi=\kappa/2\)). In (d), the rms displacement at
 \(t=0.45\)~ps is shown with a dashed vertical line whilst the solid line
 indicates the coherent part moving at the LP group velocity.  Black dashed
 curves in (a) and (c) show the result from \cref{eq:decay_ratio} with
 \(\omega^L_{k_0}\approx 0.922\)~eV, \(k_p=\pi/2\)~\(\mu\text{m}^{-1}\),
 starting from the peak at \(t=0.15\)~ps. Parameter values used for all figures
 were, in units relative to \(\omega_0^{\phantom{\dagger}}=1\)~eV,
 \(\omega_c^{\phantom{\dagger}}=0.9\), \(\Omega=0.05\) and \(\kappa=0.01\).  We
 set \(N_k=601\) positions for a chain of length \(L=200\)~\(\mu\)m and
 \(N=10^6\) molecules.
    }
    \label{fig:2}
\end{figure}

Dynamics following resonant excitation of the LP branch at \(k_p =
\pi/2~\mu\text{m}^{-1}\) are shown in \cref{fig:2}.  We examine both the photon
population \(n^{\text{ph}}\) and exciton population \(n^\text{m}\) as functions
of position, for molecular dephasing \(\gamma_\phi=0\) and
\(\gamma_\phi=\kappa/2\).

First, in the absence of dephasing  (\cref{fig:2a,fig:2b}), there is no
decoherence, and  \(n^{\text{ph}}\) is depleted solely by cavity decay
\(\kappa\),  while \(n^{\text{m}}\) exhibits a transient signal  that mirrors
\(n^{\text{ph}}\), i.e., follows the polariton.  Both populations propagate at
the  expected group velocity \(v^{\text{grp}}_{k_p}\), as may be verified by
calculating their mean or root-mean-square (rms) displacement from the initial
pump spot \(r_n=0\).

Introducing dephasing (\cref{fig:2c,fig:2d}) significantly  alters the dynamics.
The cavity population \(n^{\text{ph}}\)  depletes more rapidly, as the total
linewidth is determined  by both \(\kappa\) and \(\gamma_\phi\). In particular,
transfer  to incoherent dark exciton states occurs at rate \(\gamma_\phi\),  as
discussed further below.  

As a result, a ``dragged'' molecular population \(n^{\text{m}}\) appears from
the passing polariton wavepacket, as local incoherent excitons are created but
remain stationary~\cite{pandya2021, pandya2022}.  Crucially, when considering
both coherent \textit{and} incoherent  populations, the rms displacement
propagates with sub-group-velocity.  As the polariton signal diminishes, there
is a transition from polariton-dominated ballistic motion to a stationary (or
near-stationary) excitonic distribution.  Sub-ballistic, i.e., diffusive
behavior may be observed, even when the model does not incorporate explicit
diffusion processes.

The inclusion of exciton hopping---an inherently diffusive process---further
impedes transport through increased exciton character and group velocity
renormalization (see SI).  Conversely, we find that the presence of
exciton-exciton annihilation may lead to \textit{increased} ballistic
transport, by diminishing the excitonic tails without altering
the excitonic fraction, a surprising possibility that has also been observed in
related systems~\cite{kulig2018,sridharan2019}.

In \cref{fig:2a,fig:2c}, we also plot an analytical  prediction for the decay of
the polariton---and so photon---population with position, derived from the
mean-field equations in the linear regime (see SI): given the population
\(n^{\text{ph}}_{n_0}\) at site \(n_0\) (position \(r_{n_0}\)),
\begin{align}
    n^{\text{ph}}_n
    = 
     n^{\text{ph}}_{n_0}
    e^{\lambda(\omega^L_{k_0}) |{n-n_0}| }, \label{eq:decay_ratio}
\end{align}
where
\begin{align}
   \lambda^2(\omega)  &= -\frac{1}{C}  
   \left[(\omega- {\omega}^{\vphantom{\dagger}}_{c})  
   + i\kappa/2+\Omega^2 \chi(\omega)\right],
\end{align}
with \(\chi(\omega)\) denoting the linear molecular susceptibility. 

We note \(\lambda^2\) is proportional to the inverse retarded photon Green’s
function \(D^R(\omega)\) \cite{zeb2018}.  Hence, \(D^R(\omega)\) serves an
important figure of merit for lateral transport within a cavity, providing a
characteristic length scale over which excitations can propagate.
\cref{eq:decay_ratio} can be regarded as a version of the famous Beer-Lambert
law for lateral polariton transport \cite{mukamel1995principles}.\\

 The role of
molecular dephasing in transfer to the  dark manifold becomes evident from the
rate equations for the bright  (\(p^B\)) and dark (\(p^D\)) populations, which
follow from \cref{eq:mfb,eq:mfc} (see SI for derivation):
\begin{subequations}
\begin{align}
		\partial_t p^B_n &= - \gamma_\phi p^B_n  
        - 2\Omega \Im \Pi_n \cdot (1-2p^{\text{m}}_n), \label{eq:pB}\\
		\partial_t p^D_n &= + \gamma_\phi p^B_n  
        - 2\Omega \Im \Pi_n \cdot 2p^\text{m}_n.\label{eq:pD}
\end{align}
\end{subequations}
Here, \(p^{\text{m}}_n(t)=n^{\text{m}}_n/N_E = (1+\langle \sigma^z_n\rangle)/2\)
is the molecular population per molecule at site \(n\), i.e., the excitation
probability, and  \(\Im\Pi_n(t) = \Im(\langle a^{\vphantom{\dagger}}_{n} \rangle
\langle \sigma^+_{n}\rangle)\) represents light-matter correlations.  These
correlations act as a source in both equations,  though for \(p^D\), the
contribution is suppressed  at typical excitation densities \(p^{\text{m}}\ll
1\). 

Thus, bright states directly feed dark states at a rate  \(\gamma_\phi\),
leading to the behavior seen in \cref{fig:2d}.  Notably, phase information is
lost during scattering into dark  states, making this bulk incoherent transfer
irreversible in the absence of vacuum fluctuations (e.g., radiative pumping or
vibrationally-assisted scattering~\cite{perezsanchez2025}).  This aligns with
recent insights into molecular polariton
dynamics~\cite{perezsanchez2024b,schwennicke2024}:  transitions from dark
states back to polariton states occur  via single light-matter coupling \(g
\propto 1/\sqrt{N}\)~\cite{pino2015, ribeiro2018, martinez2019}  and are
therefore suppressed as \(N \to \infty\).\\

Let us now consider the full pump-probe experiment. Following Ref.~\cite{reitz2025nonlinear}, we expand all quantities
in pump (\(p\)) and probe (\(p'\)) strength, 
\begin{align}
\begin{gathered}
    \langle a^{\vphantom{\dagger}}_{n} \rangle =
	\sum_{a,b} \eta_p^a \eta_{p'}^b \alpha^{(a,b)}_n,\\
    \langle \sigma^-_n\rangle = 
	\sum_{a,b} \eta_p^a \eta_{p'}^b s^{(a,b)}_n,\quad 
    \langle \sigma^z_n\rangle =
	\sum_{a,b} \eta_p^a \eta_{p'}^b z^{(a,b)}_n.
    \end{gathered}\label{eq:expansion}
\end{align}
Substituting into the mean-field equations \cref{eq:mfb,eq:mfc,eq:mfa}, we
obtain a hierarchical system of  equations for the coefficients
\(\alpha^{(a,b)}_n\), \(s^{(a,b)}_n\), and \(z^{(a,b)}_n\), detailed in 
the SI. We retain terms up to order $(2,1)$, corresponding to an overall third-order nonlinear process, $\chi^{(3)}$. Assuming the probe field to be weaker than the pump, $\eta_{p'}\ll\eta_{p}$, higher-order contributions in the probe are both reduced in amplitude and, because the detected signal is selected along the probe propagation direction, generally emitted into different phase-matched directions, and may therefore be neglected (see SI for details).

We now identify an appropriate spectroscopic observable.  Since pump-probe
experiments are inherently nonlinear optical measurements, they capture
differential signals—specifically, the change in cavity output between pump-on
and pump-off conditions. Accordingly, we consider the differential transmission,
\(\Delta T = T^{p\text{-on}} - T^{p\text{-off}}\).

As explained in Ref.~\cite{reitz2025nonlinear}, the lowest-order contribution to
\(\Delta T\) arises from the combination  of the first-order probe and
second-order pump cavity fields, \(\Delta T(\omega) \sim \alpha^{(0,1)}(\omega)
\overline{\alpha}^{(2,1)}(\omega)\), where an overline denotes complex
conjugation.  Extending this to the multimode model, we define the
position-dependent differential transmission
\begin{align}
    \Delta T_n(\omega) = \eta_p^2 (\kappa^2/2)\,
	\Re[\alpha^{(0,1)}_n(\omega)\overline{\alpha}^{(2,1)}_n(\omega)],
    \label{eq:DeltaTn}
\end{align}
\noindent which scales with the intensity of the pump field and is independent of the probe field amplitude.
In words, this expression constitutes the heterodyning of the
third-order signal $\alpha_n^{(2,1)}$ by the probe $\alpha_n^{(0,1)}$. The setup
with counter-propagating pulses provides a simple way to separate out pump and
probe signals, which is a practical challenge in experiments.
\begin{figure}
    \centering
    \includegraphics[width=\linewidth]{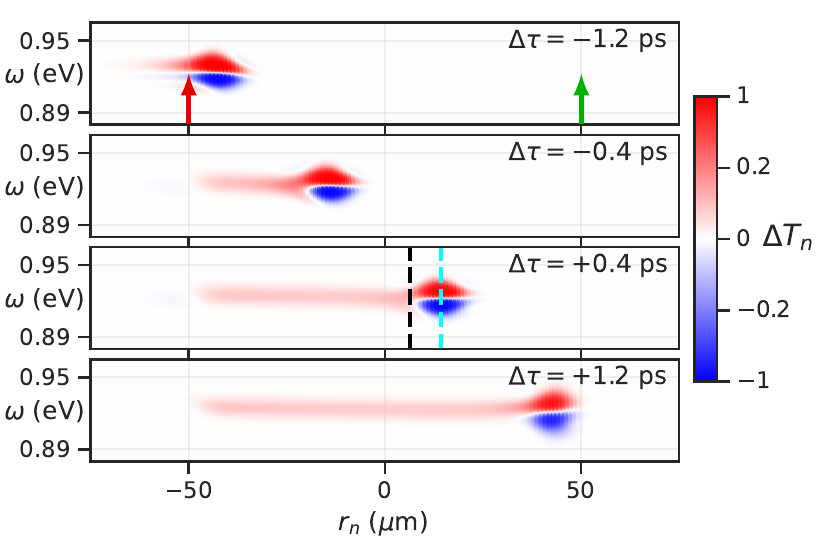}
	\vspace{-2\baselineskip}%
    \caption{
	\textit{Spatiotemporal pump-probe spectra.} Differential transmission
	\(\Delta T_n\) following evolution of a LP wavepacket. In the top panel,
	arrows indicate the spatial positions of the pump (red) and probe (green)
	spots. In the third panel, the black dashed line indicates the rms position
	of the polariton distribution, while the cyan line marks the center
	(maximum) of the polariton wavepacket.  A nonlinear colorbar scaling based
	on a \(\tanh\)-function was used to aid visibility of the trailing red
	feature.  A movie of the full dynamics is provided as Supplementary
	Information.
    }
    \label{fig:3}
\end{figure}
\Cref{fig:3} shows \(\Delta T_n(\omega)\) following pump (\(r_n=-50\)~\(\mu\)m)
and probe (\(r_n=50\)~\(\mu\)m) pulses initiated at either side of the cavity
with opposite momenta, \(k_{p}=-k_{p'}= \pi/2\)~\(\mu\text{m}^{-1}\).  The
signal is plotted for four probe delays \(\Delta \tau\), where \(\Delta \tau=0\)
corresponds to the pump and probe pulses meeting at the center of the cavity.
The direct calculation of a physical observable for imaging experiments in
\cref{eq:DeltaTn} is a main advancement of our work.

\begin{figure}
    \centering
    \includegraphics[width=\linewidth]{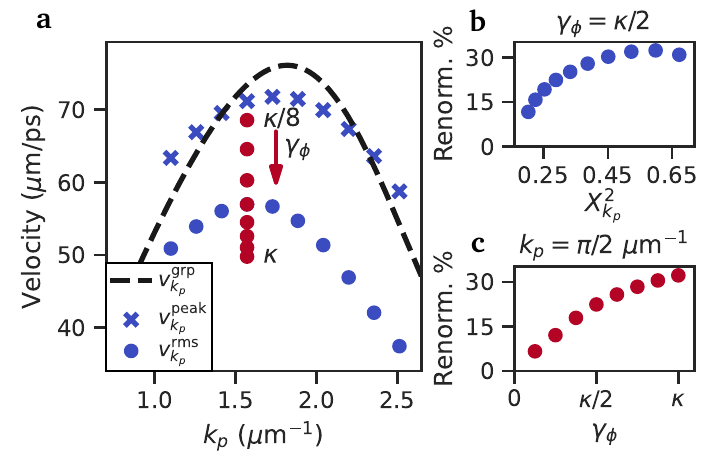}%
    \phantomsubfloat{\label{fig:4a}}%
    \phantomsubfloat{\label{fig:4b}}%
    \phantomsubfloat{\label{fig:4c}}%
	\vspace{-2\baselineskip}%
    \caption{ \textit{Transport renormalization.}
	(a) Blue: velocity of the circular feature (\(v^{\text{peak}}_{k_p}\),
	crosses) and rms displacement (\(v^{\text{rms}}_{k_p}\), circles) for
	various pump wavevectors \(k_p\) at \(\gamma_\phi=\kappa/2\). The
	velocities were calculated from
	a straight line fit to the position of the maximum and rms displacement of
	\(\abs{\Delta T_n}\) over  many delays \(-0.8\ \text{ps}\leq\Delta \tau \leq
	0.8\)~ps.  The dashed curve shows the theoretical group velocity
	\(v^{\text{grp}}_{k_p}\).
	(b) Increasing renormalization \(1- v^{\text{rms}}_{k_p}/v^{\text{grp}}_{k_p}\)
	is observed with increasing \(k_p\) and so exciton fraction \(X_{k_p}^2\).
	(a) Red: \(v^{\text{rms}}_{k_p}\) at \(k_p=\pi/2\ \mu\text{m}^{-1}\) and
	\(\gamma_\phi\) increasing from \(\kappa/8\) to \(\kappa\) in increments of
	\(\kappa/8\).  The corresponding renormalization trend is shown in (c).
    }
    \label{fig:4}
\end{figure}

The leading circular feature in each panel of \cref{fig:3} is the coherent
excitation, i.e., the polariton, with red-blue lobes resulting from Rabi
contraction  following blue shift of the LP~\cite{reitz2025nonlinear,delpo2020}
(originating from the creation of dephased population in the excited state,
corresponding to dark states \cite{xiang2021molecular}). Tracking the position
of this feature, determined approximately by the maximum of the absolute signal
(cyan line in third panel of \cref{fig:3}), we find the extracted transport
velocity (see SI for details) is close to the theoretical group velocity
\(v^{\text{grp}}_{k_p}\) from the dispersion curve (\cref{fig:4a}).  

In addition, a trailing feature in \(\Delta T_n\) emerges,  corresponding to the
population of static dark excitonic states.  Consequently, the rms displacement
of the entire  signal, which is often reported in experiments, propagates
significantly below \(v^{\text{grp}}_{k_p}\), as shown in \cref{fig:4a}.  We
note that the situation in a real experiment~\cite{xu2023} is likely more
complex than	the picture presented here, i.e., one might not observe a clear
distinction between incoherent and coherent parts of the signal.  We comment on
appropriate extensions to our model below.

Next, we repeat the simulation for different pump-probe momenta
\(\abs{k_p}=\abs{k_{p'}}\). Extracting the  rms velocity
\(v^{\text{rms}}_{k_p}\)  in each case, we find the velocity renormalization to
increase with exciton content, \cref{fig:4b}, as has been frequently reported in
experiments~\cite{rozenman2018,balasubrahmaniyam2023,berghuis2022,berghuis2024,pandya2021,
pandya2022,rozenman2018,xu2023,jin2023, cheng2023}.  We further show how
increasing \(\gamma_\phi\) at fixed pump-probe momenta also increases the
renormalization (red dots in \cref{fig:4a,fig:4c}). Both trends align with our
explanation of sub-group-velocity  transport due to stationary dark excitons,
since the feeding of these populations is set by both the molecular fraction of
the  polariton, \(p^B\), and the dephasing rate \(\gamma_\phi\).  Conversely, we
expect decreasing \(\kappa\) to protect the fraction of the propagating exciton
that is polariton, hence reduce the velocity renormalization, in line with
observed trends versus cavity $Q$-factor~\cite{pandya2022,tichauer2023}.\\

In conclusion, we developed a framework for molecular polariton transport in multimode cavities
by  calculating spatiotemporally-resolved spectra, providing a clear
understanding of how dark excitonic populations can renormalize  the transport
properties in a simplified model of a pump-probe experiment.  By analyzing the
nonlinear differential transmission, we  demonstrated how irreversible transfer
to dark exciton states,  induced by molecular dephasing, creates stationary
excitonic  populations that slow net transport.  Moreover, that this
renormalization becomes more pronounced  with increased excitonic weight
\(X_k^2\) and molecular dephasing \(\gamma_\phi\).  Similar dependence on
\(X_k^2\) has been widely reported in experiments, while that on \(\gamma_\phi\)
could be investigated, e.g., using materials with different electron-phonon
coupling strengths. 

We considered the lowest-order nonlinear signal for a counter-propagating
pump-probe experimental setup, which has the advantage of providing clear
separation of the pump and probe signals. While we used a simple model of pure
dephasing for molecular emitters, our approach readily extends to multi-level
systems, including coupling to vibrational modes directly~\cite{strashko2018}.
An accurate description of vibronic coupling may be essential for a complete
understanding of real experiments where these interactions dominate the
dynamics~\cite{xu2023}.

Furthermore, this method could be applied to calculate higher-order nonlinear
responses and explore alternative experimental schemes not limited to
microcavity exciton-polaritons.  For example, one may consider setups involving
local probe fields that are scanned to construct a real-space
image~\cite{jin2023}.

Future work may also incorporate higher-order correlations beyond the mean-field
description to understand analogous microscopy experiments based on
photoluminescence measurements~\cite{sanchez-barquilla2020}, or integrate our
approach with FDTD  electromagnetic simulations~\cite{zhou2024b,zhou2024}  to
account for arbitrary photonic environments. 

\section*{Acknowledgements}%
This research was primarily supported by the Air Force Office of Scientific
Research (AFOSR) through the Multi-University Research Initiative (MURI) program
no.~FA9550-22-1-0317.  We acknowledge helpful discussions with Milan Delor,
Yongseok Hong and Harsh Bhakta.  P.~F.-W. thanks Jonathan Keeling, Brendon
Lovett and Niya Petkova for formative discussions. 

\bibliography{refs.bib}

\clearpage
\onecolumngrid

\setcounter{section}{0}
\setcounter{equation}{0}
\setcounter{figure}{0}
\setcounter{table}{0}

\renewcommand{\thesection}{S\arabic{section}}
\renewcommand{\theequation}{S\arabic{equation}}
\renewcommand{\thefigure}{S\arabic{figure}}
\renewcommand{\thetable}{S\arabic{table}}

\begin{center}
\large
Supplementary Information for\\[4pt]
\textbf{``Mapping molecular polariton transport via pump-probe microscopy''}
\end{center}

\phantomsection
\vspace{0.5em}
\begin{center}
\large\bfseries Contents
\end{center}
\vspace{0.25em}

{
\setcounter{tocdepth}{2}

% tighten TOC spacing
\parskip=0pt
\itemsep=0pt
\parsep=0pt

\makeatletter
% reduce vertical skip before section/subsection entries in the TOC
\def\l@section#1#2{\addpenalty\@secpenalty\addvspace{0.2em}%
  \setlength\@tempdima{1.5em}%
  \begingroup
    \parindent \z@ \rightskip \@pnumwidth
    \parfillskip -\@pnumwidth
    \leavevmode
    \advance\leftskip\@tempdima
    \hskip -\leftskip
    #1\nobreak\hfil \nobreak\hb@xt@\@pnumwidth{\hss #2}\par
  \endgroup
}

\def\l@subsection#1#2{%
  \setlength\@tempdima{2.8em}%
  \begingroup
    \parindent \z@ \rightskip \@pnumwidth
    \parfillskip -\@pnumwidth
    \leavevmode
    \advance\leftskip\@tempdima
    \hskip -\leftskip
    #1\nobreak\hfil \nobreak\hb@xt@\@pnumwidth{\hss #2}\par
  \endgroup
}

\addtocontents{\supplementarytocfile}{\protect\setcounter{tocdepth}{2}}
\@starttoc{\supplementarytocfile}
\makeatother
}

\clearpage
\twocolumngrid

\section{List of parameters}
\begin{table}[h]\centering
\begin{tabular}{c|c|c}
\textbf{Symbol} & \textbf{Description} & \textbf{Value/Def.}\textsuperscript{*}  \\\hline
\(N\) & Number of molecules & \(10^6\) \\
\(N_k\) & Number of cavity modes & \(601\) \\
\(L\) & System length & \(200\ \mu\text{m}\) \\
\(N_s\) & Number of molecules/site & $N/N_k$ \\
\(\Delta r\) & Lattice spacing \(\Delta r = L/N_k\) & \(0.33\ \mu\text{m}\) \\
\(\omega_0^{\phantom{\dagger}}\) & Molecule transition frequency & \(1.0\ \text{eV}\) \\
\(\omega_c^{\phantom{\dagger}}\) & Cavity frequency at \(k=0\) & \(0.9\ \text{eV}\) \\
$C$ & photon propagation speed & $c^2/(2\omega_c\Delta r)$\\
\(g\) & Individual light-matter coupling & \(\Omega/\sqrt{N}\) \\
\(\Omega\) & Collective light-matter coupling & \(0.05\ \text{eV}\) \\
\(\kappa\) & Photon decay rate & \(0.01\ \text{eV}\) \\
\(\gamma_\phi\) & Molecular dephasing rate & \(0.005\ \text{eV}\)\textsuperscript{**} \\
\(\sigma_t\) & Pulse width (time) & \(25\ \text{fs}\) \\
\(\sigma_r\) & Pulse width (space) & \(5\ \mu\text{m}\) \\
\(k_{p/p'}\) & Pump/probe pulse wavevector & \(\pm\pi/2\ \mu\text{m}^{-1}\)\textsuperscript{\(\dagger\)} \\
\(\Delta \tau\) & Pump-probe delay & \(-1.2\leq \Delta \tau \leq 1.2\ \text{ps}\)\textsuperscript{\(\ddagger\)} \\
\end{tabular}
\caption{%
Main parameters used in the Letter and Supporting Information.
\vspace{4pt}\\
\footnotesize
\({}^{\ast}\) Note that, when defining the model in the Letter, we set \(\hbar=1\), such that the units of energy coincide with that of inverse time via \(E=\hbar \omega\). The conversion factor to obtain the chosen physical units of electron volts (eV) is \((\hbar/e)\).\\
\({}^{\ast\ast}\) Varies in Fig.~4c.\\
\({}^\dagger\) Varies in Figs.~4a,b.\\
\({}^\ddagger\) Varies in Fig.~3.%
}
\label{table:parameters}
\end{table}

\section{Derivation of mean-field equations}
In this section, we outline the derivation of the mean-field equations of motion
for the model introduced in Eq.~(3) of the Letter. These steps may be applied to
other many-body models with unitary evolution and incoherent processes to
calculate spatially resolved dynamics in extended systems.

For reference, we repeat here
 the TC Hamiltonian introduced in Eq.~(1)  
\begin{align}
\begin{split}
	H_{} &= \sum_{k=1}^{N_k}
	{\omega}^{\vphantom{\dagger}}_{k} a^{\dagger}_{k}a^{\vphantom{\dagger}}_{k}
    +  \sum_{n=1}^{N} \frac{\omega_0^{\vphantom{\dagger}}}{2} \sigma^z_{n}\\ &+   \sum_{k=1}^{N_k}\sum_{n=1}^{N}
	g(a^{\vphantom{\dagger}}_{k}\sigma^+_{n}e^{ikr_n} + 
    \text{H.c.}),
\end{split}
\label{sm:eq:H}
\end{align}
and Master Equation
[Eq. (3) in the main text]:
\begin{align}
		\partial_t \rho = &-i \bigl[ H , \rho \bigr]
	+ \sum_{k=1}^{N_k} \kappa \mathcal{L}[a^{\vphantom{\dagger}}_{k}]
    + \sum_{n=1}^{N} \frac{\gamma_\phi}{4}  \mathcal{L}[\sigma^z_n].
	\label{am:eq:ME}
\end{align}

\subsection{Spatial Coarse Graining}

The first step is to coarse grain the one-dimensional
molecular system by assigning molecules into sites at
\(r_n=\Delta r, 2 \Delta r, \ldots , N_k \Delta r\)
such that the spatial resolution \(\Delta r = L/N_k\) is
set by the number of sites \(N_k\). For this purpose, it
is helpful to introduce a two-index notation for the spin
variables (Pauli matrices),
\(\sigma^\alpha_{n\x}\) where \(n=1,\ldots, N_k\)
identifies the site and \(\x=1,\ldots,N_s\) the molecule 
at that site. Here \(N_s=N/N_k\) is the number of molecules
per site:
\begin{align}
	\begin{split}
	H_{} &= \sum_{k=1}^{N_k}
	{\omega}^{\vphantom{\dagger}}_{k} a^{\dagger}_{k}a^{\vphantom{\dagger}}_{k}
	+ \sum_{n=1}^{N_k} \sum_{\x=1}^{N_s}  
	\frac{\omega_0}{2} \sigma^z_{n\x} \\
    &+  \sum_{k=1}^{N_k}\sum_{n=1}^{N_k}\sum_{\x=1}^{N_s} 
	g(a^{\vphantom{\dagger}}_{k}\sigma^+_{n\x}e^{ikr_n} + 
    \text{H.c.})
	,\end{split}
\label{sm:eq:H2}
\end{align}
\begin{align}
		\partial_t \rho = &-i \bigl[ H , \rho \bigr]
	+ \sum_{k=1}^{N_k} \kappa \mathcal{L}[a^{\vphantom{\dagger}}_{k}]
    + \sum_{n=1}^{N_k} \sum_{\x=1}^{N_s} \frac{\gamma_\phi}{4}\mathcal{L}[\sigma^z_{n\x}].
	\label{sm:eq:ME2b}
\end{align}

\subsection{Mean-field Approximation}
We next write down equations of motion for the operators 
\(a^{\phantom{\dagger}}_k\), \(\sigma^\alpha_{n\x}\). Later,
we will realize that we only need to keep a single representative
spin variable for \(n=1\),\ldots, \(N_k\), since molecules
at the same site have the same on-site properties 
(i.e., operator expectation values).

The Heisenberg equations of motion for the expectation
of an operator \(X\) is obtained from \cref{sm:eq:ME2b} 
via \(\partial_t \langle X \rangle = \text{Tr}[X \partial_t \rho]\). This yields
\begin{subequations}
\begin{align}
\begin{split}
    \partial_t \langle a^{\phantom{\dagger}}_k \rangle
    &=  	-\bigl( i {\omega}^{\vphantom{\dagger}}_{k} + \kappa/2 \bigr) \langle
		a^{\vphantom{\dagger}}_{k} \rangle\\
&- i g \sum_{n=1}^{N_k} \sum_{\x=1}^{N_s} e^{-ikr_n} \langle \sigma^-_{n\x} \rangle ,
\end{split} \label{sm:eq:mf1a} \\
\begin{split}
\partial_t \langle \sigma_{n\x}^-\rangle
&= -\bigl( i \omega^{\phantom{\dagger}}_0 +  \gamma_\phi/2 \bigr) \langle \sigma_{n\x}^- \rangle \\
&+i g \sum_{k=1}^{N_k}
e^{ikr_n}  \langle  \sigma^z_{n\x} a^{\phantom{\dagger}}_k \rangle,
\end{split} \label{sm:eq:mf1b} \\
\partial_t \langle \sigma_{n\x}^z\rangle
&= -4g \sum_{k=1}^{N_k}\Im[\langle \sigma_{n\x}^-  a^{\dagger}_k  e^{-ikr_n} \rangle ]  .
\label{sm:eq:mf1c}
\end{align}
\end{subequations}
Note that sums over the spin variables do not appear in 
\crefrange{sm:eq:mf1b}{sm:eq:mf1c}, since 
\(\sigma^\alpha_{n\x}\) refers to a single species
and so commutes with \(\sigma^\alpha_{n\y}\) 
unless \(m=n\) \textit{and} \(\y=\x\). Additionally,
these equations contain expectations of pairs of operators
and so \crefrange{sm:eq:mf1a}{sm:eq:mf1c} does not form
a closed set of equations.

To close \crefrange{sm:eq:mf1a}{sm:eq:mf1c}, we apply the 
mean-field approximation by factorizing the expectations
\( \langle  \sigma^z_{n\x} a^{\phantom{\dagger}}_k \rangle
\approx  \langle  \sigma^z_{n\x} \rangle \langle a^{\phantom{\dagger}}_k \rangle\),
\(\langle \sigma_{n\x}^-  a^{\dagger}_k\rangle
\approx 
\langle \sigma_{n\x}^-\rangle \langle  
a^{\phantom{\dagger}}_k\rangle^*\). As explained in
the Letter, this follows directly from a product
state ansatz for the total many-body state. Specifically,
it is assumed
\begin{align}
    \rho = \bigotimes_{k=1}^{N_k} \rho_{k} \bigotimes_{n=1}^{N_k}\bigotimes_{\x=1}^{N_s} \rho_{n\x},
\end{align}
where \(\rho_{k}\) is a density matrix for the 
\(k^{\text{th}}\) photon mode and \(\rho_{n\x}\) are the
density matrices for each molecule. The equations become
\begin{subequations}
\begin{align}
\begin{split}
    \partial_t \langle a^{\phantom{\dagger}}_k \rangle
    &=  	-\bigl( i {\omega}^{\vphantom{\dagger}}_{k} + \kappa/2 \bigr) \langle
		a^{\vphantom{\dagger}}_{k} \rangle\\
&- i g \sum_{n=1}^{N_k} \sum_{\x=1}^{N_s} e^{-ikr_n} \langle \sigma^-_{n\x} \rangle ,
\end{split} \label{sm:eq:mf2a} \\
\begin{split}
\partial_t \langle \sigma_{n\x}^-\rangle
&= -\bigl( i \omega^{\phantom{\dagger}}_0 +  \gamma_\phi/2 \bigr) \langle \sigma_{n\x}^- \rangle \\
&+i g \langle \sigma^z_{n\x} \rangle \sum_{k=1}^{N_k}
e^{ikr_n}  \langle   a^{\phantom{\dagger}}_k \rangle,
\end{split} \label{sm:eq:mf2b} \\
\partial_t \langle \sigma_{n\x}^z\rangle
&=  -4g\sum_{k=1}^{N_k}\Im[\langle \sigma_{n\x}^-  \rangle \langle a^{\phantom{\dagger}}_k\rangle^*
 e^{-ikr_n}]  .
\label{sm:eq:mf2c}
\end{align}
\end{subequations}

Finally, since
the on-site properties \(\langle \sigma_{n\x}^\alpha\rangle\)
are the same for any \(\x=1,\ldots, N_s\) at fixed
\(n\) (i.e., molecules at the same location), we can
perform the sum over \(\x\) in \cref{sm:eq:mf2a} and write
\begin{subequations}
\begin{align}
\begin{split}
    \partial_t \langle a^{\phantom{\dagger}}_k \rangle
    &=  	-\bigl( i {\omega}^{\vphantom{\dagger}}_{k} + \kappa/2 \bigr) \langle
		a^{\vphantom{\dagger}}_{k} \rangle\\
&- i g N_s\sum_{n=1}^{N_k} e^{-ikr_n} \langle \sigma^-_{n} \rangle ,
\end{split} \label{sm:eq:mf3a} \\
\begin{split}
\partial_t \langle \sigma_{n}^-\rangle
&= -\bigl( i \omega^{\phantom{\dagger}}_0 +  \gamma_\phi/2 \bigr) \langle \sigma_{n}^- \rangle \\
&+i g \langle \sigma^z_{n} \rangle \sum_{k=1}^{N_k}
e^{ikr_n}  \langle   a^{\phantom{\dagger}}_k \rangle,
\end{split} \label{sm:eq:mf3b} \\
\partial_t \langle \sigma_{n}^z\rangle
&=  -4g\sum_{k=1}^{N_k}\Im[\langle \sigma_{n}^-  \rangle \langle a^{\phantom{\dagger}}_k\rangle^*
 e^{-ikr_n}],
\label{sm:eq:mf3c}
\end{align}
\end{subequations}
where now \(\langle \sigma_n^\alpha \rangle\) denotes
the expectation for \textit{any one} of the 
identical spin variables at \(r_n\).

\subsection{Photonic Variables in Real Space}
It is convenient to evaluate the photon variables in
real space by introducing the transformed
operators,
\begin{align}
    a^{\vphantom{\dagger}}_{n}=\frac{1}{\sqrt{N_k}} \sum_{k=1}^{N_k}
e^{ikr_n}a^{\vphantom{\dagger}}_{k},
\end{align}
from which
\begin{subequations}
\begin{align}
\begin{split}
    \partial_t \langle a^{\phantom{\dagger}}_n \rangle
    &=  	- \frac{i}{\sqrt{N_k}} \sum_k 
    e^{ikr_n} {\omega}^{\vphantom{\dagger}}_{k}
     \langle
		a^{\vphantom{\dagger}}_{k} \rangle
- \bigl( \kappa/2 \bigr) \langle
		a^{\vphantom{\dagger}}_{n} \rangle\\
&- i g N_s\sqrt{N_k} \langle \sigma^-_{n} \rangle ,
\end{split} \label{sm:eq:mf4a} \\
\begin{split}
\partial_t \langle \sigma_{n}^-\rangle
&= -\bigl( i \omega^{\phantom{\dagger}}_0 +  \gamma_\phi/2 \bigr) \langle \sigma_{n}^- \rangle \\
&+i g \sqrt{N_k}\langle \sigma^z_{n} \rangle 
 \langle   a^{\phantom{\dagger}}_n \rangle,
\end{split} \label{sm:eq:mf4b} \\
\partial_t \langle \sigma_{n}^z\rangle
&=  -4g\sqrt{N_k} 
\Im[\langle \sigma_{n}^-  \rangle \langle 
a^{\phantom{\dagger}}_n\rangle^*],
\label{sm:eq:mf4c}
\end{align}
\end{subequations}
having used 
\(\sum_{m=1}^{N_k}\sum_{k=1}^{N_k} e^{-ikr_m}=N_k\delta_{nm}\)
to obtain the second line in \cref{sm:eq:mf4a}. To resolve
the remaining sum in this equation, we note that, for
%a quadratic dispersion
\( {\omega}^{\vphantom{\dagger}}_{k} = 
{\omega}^{\vphantom{\dagger}}_{c}
+ k^2c^2/2 {\omega}^{\vphantom{\dagger}}_{c}\),
\begin{align}
\begin{split}
   \frac{1}{\sqrt{N_k}}& \sum_{k=1}^{N_k} e^{ikr_n}{\omega}^{\vphantom{\dagger}}_{k} 
    a^{\vphantom{\dagger}}_{k}
    =
    \frac{1}{\sqrt{N_k}}  {\omega}^{\vphantom{\dagger}}_{c} \sum_{k=1}^{N_k} e^{ikr_n}
     a^{\vphantom{\dagger}}_{k}\\\nonumber
    & +
     \frac{1}{\sqrt{N_k}}
    \frac{c^2}{2 {\omega}^{\vphantom{\dagger}}_{c}
    }\sum_{k=1}^{N_k} e^{ikr_n}k^2
     a^{\vphantom{\dagger}}_{k} 
     \end{split}\\
     &= 
      {\omega}^{\vphantom{\dagger}}_{c} 
     a^{\vphantom{\dagger}}_{n} +
     \frac{1}{\sqrt{N_k}}
    \frac{c^2}{2 {\omega}^{\vphantom{\dagger}}_{c}
    }
     \partial_{r_n}^2
     \left(\sum_{k=1}^{N_k}  
     e^{ikr_n}
     a^{\vphantom{\dagger}}_{k}\right) \\\nonumber
          &= 
            {\omega}^{\vphantom{\dagger}}_{c} 
     a^{\vphantom{\dagger}}_{n} +
     \frac{1}{\sqrt{N_k}}
    \frac{c^2}{2 {\omega}^{\vphantom{\dagger}}_{c}
    \Delta r^2}\partial_{n}^2
     \left(\sum_{k=1}^{N_k}  
     e^{ikr_n}
     a^{\vphantom{\dagger}}_{k}\right) \\\nonumber
     &= 
     {\omega}^{\vphantom{\dagger}}_{c} 
     a^{\vphantom{\dagger}}_{n} +
     \frac{c^2}{{\omega}^{\vphantom{\dagger}}_{c} \Delta r^2} \partial_n^2 a^{\vphantom{\dagger}}_{n},
\end{align}
where the second last line follows from the chain rule.
Hence a quadratic dispersion maps to a second spatial
derivative, and the equations in real space read
\begin{subequations}
\begin{align}
\begin{split}
    \partial_t \langle a^{\phantom{\dagger}}_n \rangle
    &=  	- \bigl(i \omega^{\phantom{\dagger}}_c + \kappa/2\bigr)\langle
		a^{\vphantom{\dagger}}_{n} \rangle
        - i g N_s\sqrt{N_k} \langle \sigma^-_{n} \rangle \\
&+ \frac{ic^2}{2 \omega^{\phantom{\dagger}}_c \Delta r^2}
\langle
		a^{\vphantom{\dagger}}_{n} \rangle,
\end{split} \label{sm:eq:mf6a} \\
\begin{split}
\partial_t \langle \sigma_{n}^-\rangle
&= -\bigl( i \omega^{\phantom{\dagger}}_0 +  \gamma_\phi/2 \bigr) \langle \sigma_{n}^- \rangle \\
&+i g \sqrt{N_k}\langle \sigma^z_{n} \rangle 
 \langle   a^{\phantom{\dagger}}_n \rangle,
\end{split} \label{sm:eq:mf6b} \\
\partial_t \langle \sigma_{n}^z\rangle
&=  -4g\sqrt{N_k} 
\Im[\langle \sigma_{n}^-  \rangle \langle 
a^{\phantom{\dagger}}_n\rangle^*].
\label{sm:eq:mf6c}
\end{align}
\end{subequations}
To obtain Eqs.~(4a)-(4c) in the Letter
it remains to identify
\(\Omega=g\sqrt{N}\), \(C=c^2/2 \omega^{\phantom{\dagger}}_c\)
and use the rescaled photon variable
\(\tilde{a}_n^{\phantom{\dagger}} = \sqrt{N_k/N}a^{\phantom{\dagger}}_n \):
%\subsection{Perturbation Expansion} % include in next section

\begin{subequations}
\label{eq:mf}
\begin{align}
\begin{split}
    		\partial_t \langle \tilde{a}^{\vphantom{\dagger}}_{n}\rangle  &=
- \bigl(i {\omega}^{\vphantom{\dagger}}_{c}+ \kappa/2\bigr)
\langle \tilde{a}^{\vphantom{\dagger}}_n \rangle  
- i \Omega  \langle \sigma^-_{n} \rangle  \\
&\phantom{=}\
+iC \partial_n^2 \langle \tilde{a}^{\vphantom{\dagger}}_{n} \rangle
+ \sqrt{\kappa} \langle \tilde{a}_n^{\text{in}} \rangle,
\end{split}\label{eq:mfa}\\
		\partial_t \langle \sigma^-_{n} \rangle  &= -
		\bigl( i \omega_0^{\vphantom{\dagger}} +
\gamma_\phi/2 \bigr) \langle \sigma^-_{n} \rangle
+ i \Omega
\langle \sigma^z_{n} \rangle \langle \tilde{a}^{\vphantom{\dagger}}_{n} \rangle ,
		\label{eq:mfb}	
		\\
		\partial_t \langle \sigma^z_{n} \rangle &= 
		- 4 \Omega
		\Im \left[ \langle \sigma^-_{n} \rangle\langle
				\tilde{a}^{\vphantom{\dagger}}_{n}\rangle^*
		\right].\label{eq:mfc}
\end{align}
\end{subequations}

\section{Polaritonic Beer-Lambert law}
To obtain an analytical expression for the  propagation of photonic modes in the
cavity,  we linearize the mean-field equations by writing  \(\langle \sigma^z_n
\rangle = z_n^{(0,0)} + \delta z_n\),  where \(z_n^{(0,0)}\) is
time-independent.  In a leading-order approximation, we assume that changes  to
the molecular population are small,  \(|\delta z_n| \ll |z_n^{(0,0)}|\),  such
that \cref{eq:mfa,eq:mfb}
for the photon and  coherence expectations
form a closed set:  
\begin{subequations}
\begin{align}
\begin{split}
        		\partial_t \langle \tilde{a}^{\vphantom{\dagger}}_{n}\rangle  &=
- (i {\omega}^{\vphantom{\dagger}}_{c}+ \kappa/2)
\langle \tilde{a}^{\vphantom{\dagger}}_n \rangle  
- i \Omega  \langle \sigma^-_{n} \rangle  
 \\
&\phantom{=}\
+iC \partial_n^2 \langle \tilde{a}^{\vphantom{\dagger}}_{n} \rangle,
\label{eq:mf_lineara}
\end{split}\\
		\partial_t \langle \sigma^-_{n} \rangle  &= - ( i  \omega_0^{\vphantom{\dagger}} +
\gamma_\phi/2 ) \langle \sigma^-_{n} \rangle
+ i \Omega z^{(0,0)}_n\langle \tilde{a}^{\vphantom{\dagger}}_{n} \rangle.
\label{eq:mf_linearb}
\end{align}
\end{subequations}
For simplicity we set \(\langle \tilde{a}^{\text{in}}_n\rangle =0\), to derive
an expression after the driving fields have switched off, but the following is
readily generalized to include non-zero input. 

\Cref{eq:mf_lineara,eq:mf_linearb} can be solved by taking Fourier transforms.
Defining \(\alpha_n^{\vphantom{\dagger}}(\omega)  = \int dt\, e^{i \omega t}
\langle \tilde{a}^{\vphantom{\dagger}}_{n} \rangle(t)\), \(s_n(\omega)  = \int
dt\, e^{i \omega t} \langle {\sigma}^{-}_{n} \rangle(t)\), we find
\begin{align}
    -i \omega s_n(\omega) &= -
     ( i \omega_0^{\vphantom{\dagger}} +
\gamma_\phi/2 )  s_{n} (\omega)
+ i \Omega z^{(0,0)}_n \alpha_n^{\vphantom{\dagger}}(\omega) \\
\Rightarrow s_n(\omega) &= - \Omega \chi(\omega)
\alpha_n^{\vphantom{\dagger}}(\omega), \label{eq:sigma_omega}
\end{align}
where the molecular susceptibility is defined as
\begin{align}
    \chi(\omega) =
    \frac{z_n^{(0,0)}}{(\omega-  \omega_0^{\vphantom{\dagger}})+i \gamma_\phi/2}.
\end{align}
For a uniform molecular population (e.g., all molecules in the ground state,  
\(z_n^{(0,0)} \equiv -1\)), \(\chi(\omega)\) is independent of position \(n\). 

Substituting \cref{eq:sigma_omega} into the equation for
\(\alpha_n^{\vphantom{\dagger}}(\omega)\)  yields
\begin{align}
\begin{split}
    -i\omega \alpha_n^{\vphantom{\dagger}}(\omega) &= 
    - (i {\omega}^{\vphantom{\dagger}}_{c}+ \kappa/2) \alpha_n^{\vphantom{\dagger}}(\omega)
    + i C \partial_n^2 \alpha_n^{\vphantom{\dagger}}(\omega)\\
   &\phantom{=}\ -i \Omega s_n(\omega)
    \end{split} \\
    \Rightarrow \partial_n^2 \alpha_n^{\vphantom{\dagger}}(\omega) &= \lambda^2(\omega)
    \alpha_n^{\vphantom{\dagger}}(\omega),\label{eq:Helmholtz}
\end{align}
where \(\lambda^2(\omega) = -(1/C) \left[
(\omega-{\omega}^{\vphantom{\dagger}}_{c}) + i \kappa/2 + \Omega^2\chi(\omega)
\right] \) relates to the retarded photon Green's function, as described in the
Letter.

\Cref{eq:Helmholtz} is a (1D) Helmholtz equation, which arises commonly, e.g.,
in electromagnetism~\cite{novotny2012} and quantum scattering
problems~\cite{shankar1994}.  It admits two solutions:
\begin{align}
    \alpha_n^{\rightarrow}(\omega) = Ae^{\lambda(\omega) n},\quad
    \alpha_n^{\leftarrow}(\omega) = Ae^{-\lambda(\omega) n},
\end{align}
corresponding to right- and left-propagating waves when \(\Im \lambda(\omega) >
0\).  Accordingly, the real part of the propagation constant, \(\Re
\lambda(\omega) = \Re[D^R(\omega)]^{-1/2}\)  governs the decay of the
intracavity field at frequency \(\omega\),  as expressed in Eq.~(6) of the Letter.
We note further that, whilst derived for our simplified
model comprising two-level systems, this result can be naturally generalized to
complex multi-level systems where the susceptibility \(\chi(\omega)\) is
replaced by the full molecular self-energy~\cite{yuen-zhou2024}.

\section{Bright and dark exciton populations}
Bright excitonic states are symmetric superpositions of all  emitter states that
couple to a cavity mode~\cite{zeb2022}.  From the interaction in \cref{sm:eq:H}, we
define the bright-state  operator for the \(k^{\text{th}}\) mode as: 
\begin{align}
    \sigma^+_k = \frac{1}{\sqrt{N}} \sum_{n=1}^{N} e^{-ikr_n} \sigma^+_n.
\end{align}
Applying our spatial coarse-graining to \(N_k\) positions,  we rewrite the sum
over all molecules as a sum over positions  \(r_n = n\Delta r \)
(\(n=1,\ldots,N_k\)) and the  \(N_s\) molecules at each position:  
\begin{align}
    \sigma^+_k = \frac{1}{\sqrt{N}} \sum_{n=1}^{N_k}\sum_{\x=1}^{\text{\(N_s\)}}
    e^{-ikr_n} \sigma^+_{n\x}.
\end{align}
The total bright excitonic population as a function of position  is then
obtained from  
\begin{align}
    p^{B\text{-tot}}_n = \frac{1}{N_k} \sum_{k, k'} \langle \sigma^+_{k} \sigma^-_{k'} \rangle e^{i(k'-k)r_n}.\label{eq:pb_def}
\end{align}

When evaluating \cref{eq:pb_def}, care must be taken when calculating
coherences \(\langle \sigma^+_{n\x}\sigma^-_{m\y} \rangle\) for  \(n=m\) and
\(\x=\y\), i.e., when the indices refer to the same molecule.  In this case,
\(\langle \sigma^+_{n\x}\sigma^-_{n\x} \rangle \equiv  (1+\langle \sigma^z_{n\x}
\rangle)/2\), and a mean-field approximation  should not be applied since the operators are acting on the same molecule.  With this
in mind, from \cref{eq:pb_def} it follows that
\begin{align}
    p^{B\text{-tot}}_n &= (\text{\(N_s\)}-1) |\langle\sigma^+_{n\x}\rangle|^2 +
	\frac{1}{2}\left( \langle \sigma^z_{n\x} \rangle		+1\right).
\end{align}
Since molecules at the same position share identical on-site properties,  we
write \(\langle \sigma_{n\x}^+ \rangle = \langle \sigma_n^+ \rangle\),  where
\(\langle \sigma_n^+ \rangle\) is the expectation for any molecule at \(n\).  In
the large-\(N\) limit, where \(\text{\(N_s\)} = N/N_k \gg 1\), we obtain
\begin{align}
   p^{B\text{-tot}}_n &\approx N_s|\langle\sigma^+_{n}\rangle|^2 .
\end{align}
The dark state population \(p^{D\text{-tot}}_n\) is given by the total  
molecular population at \(n\),  
\(p^{m\text{-tot}}_n = \text{\(N_s\)} (1 + \langle \sigma^z_n \rangle)\),  
minus the bright-state contribution: 
\begin{align}
\nonumber    p^{D\text{-tot}}_n &= p^{m\text{-tot}}_n - p^{B\text{-tot}}_n \\
    &\approx 
    \text{\(N_s\)}
		\left[ \left( 1 + \langle \sigma^z_{n} \rangle \right)/2 -
		|\langle\sigma^+_{n}\rangle|^2\right].
\end{align}
Eq.~(8a) and (8b) of the Letter for the dynamics of the scaled populations
\(p^B_n=p^{B\text{-tot}}_n/N_E\) and \(p^D_n=p^{D\text{-tot}}_n/\text{\(N_s\)}\) follow
straightforwardly from the mean-field \cref{eq:mfb,eq:mfc} for
\(\langle \sigma^-_n \rangle\) and \(\langle \sigma^z_n\rangle\).

\section{Derivation of Spectroscopic Observables}
In this section, we provide steps for calculating spectroscopic 
observables within the mean-field approach. 
We focus on the differential transmission in a pump-probe 
experiment as considered in the Letter, but our framework may
be applied in principle to determine linear and non-linear 
observables in pump-probe spectroscopy configurations 
in cavity-QED systems.

The derivation proceeds in two main steps, 
firstly establishing the perturbative treatment of multiple input 
pulses, then connecting the resulting field amplitudes to 
measurable observables.

\subsection{Input Pulses and Perturbative Expansion}
Dynamical information in a non-linear spectroscopic experiment is
obtained by interacting the system with \textit{multiple} light pulses.
The fact that there is more than one input pulse means that 
it is not possible to think simply in terms of evolution from 
a fixed initial state, which is the most common approach when obtaining quantum
dynamics in theoretical calculations. Instead, one must consider
these pulses as driving terms of the photonic (cavity) field in the 
evolution, hence adding the input field  % represented by 
\(\langle \tilde{a}^{\text{in}}_n (t) \rangle\) to
\cref{sm:eq:mf6a},
% N.B. sqrt{kappa} convention
\begin{align}
\begin{split}
    		\partial_t \langle \tilde{a}^{\vphantom{\dagger}}_{n}\rangle  &=
- \bigl(i {\omega}^{\vphantom{\dagger}}_{c}+ \kappa/2\bigr)
\langle \tilde{a}^{\vphantom{\dagger}}_n \rangle  
- i \Omega  \langle \sigma^-_{n} \rangle  \\
&\phantom{=}\
+iC \partial_n^2 \langle \tilde{a}^{\vphantom{\dagger}}_{n} \rangle
+ \sqrt{\kappa} \langle \tilde{a}_n^{\text{in}} \rangle,
\end{split}\label{sm:eq:mf7a}
\end{align}
which is Eq.~(4a) of the Letter. 

For a pump-probe experiment, we consider two input pulses. 
Each input pulse (\(\beta\))
is characterized by its magnitude, \(\eta_\beta\), phase
\(e^{-i \omega^{\phantom{\dagger}}_\beta t}\) 
(frequency \(\omega^{\phantom{\dagger}}_\beta\)),
and shape in both time \(f_\beta(t)\) \textit{and}
space \(D_n^\beta\) (equivalently, time and momentum space).
Writing these components in setting
\begin{align}
    \sqrt{\kappa} \langle \tilde{a}^{\text{in}}_n (t) \rangle=
    \eta_{p} f_{p}(t)  
      e^{-i  {\omega}^{\vphantom{\dagger}}_{p} t}D_n^{p} 
      +
     \eta_{p'}  f_{p'}(t)  
      e^{-i  {\omega}^{\vphantom{\dagger}}_{p'} t} 
      D_n^{p'},\label{sm:eq:ain}
\end{align}
%which is \cref{eq:ain} of the Letter, 
allows one to specify 
two pulses---i.e., pump (\(p\)) and probe (\(p'\))---of
any possible shape. In the Letter we consider 
Gaussian pulses in the form
\begin{subequations}
\begin{align}
    f_\beta(t)e^{-i\omega_{\beta}^{\phantom{\dagger}}t} &= \frac{1}{\sqrt{2\pi \sigma_t^2}} e^{-(t-t_0)^2/(2\sigma_t^2)}e^{-i\omega_{k_\beta}^{\text{LP}}t},  \\
    D_n^\beta &= \mathcal{N} e^{-(n-n_0)^2/(2\sigma_n^2)}e^{i k_\beta n \Delta r},
\end{align}
\end{subequations}
% i k_p r = i k_p Delta r n = i (2pi/L) K Delta r n = (2pi/N_k) (K Delta r) n
where the normalization \(\mathcal{N}\) is chosen such that
\(\sum_n \abs*{D_n^\beta}^2 = 1\) % = \sum_k \abs{D_k}^2
and \(\sigma_n=\sigma_r/\Delta r\) defines the pulse width in space.
As can be seen, information of the target input wavevector \(k_\beta\)
is imprinted on the spatial profile 
\(D_n^\beta\) as a modulation \( e^{i k_\beta r_n}\).

% Note that the specification of the spatial profile \(D_n^{\beta}\) is equivalent
% to specification of the profile in momentum space, \(D_k^{\beta}\). In particular
% information of the target input wavevector \(k_\beta\) is imprinted on 
% \(D_n^\beta\) as a modulation \(\sim e^{i k_\beta r_n}\).

% The central assumption...
The key insight underlying our method is that these input pulses are nominally 
\textit{weak} in the sense that one may consider them as
perturbations to the system's normal state 
\(\langle a^{\phantom{\dagger}}_n\rangle= \langle \sigma^-_n \rangle=0\),
\(\langle \sigma^z_n \rangle = -1\), i.e., an empty cavity. This permits 
a perturbative expansion of the system state in powers of
the pulse magnitudes,
\begin{align}
\begin{gathered}
    \langle a^{\vphantom{\dagger}}_{n} \rangle =
	\sum_{a,b} \eta_p^a \eta_{p'}^b \alpha^{(a,b)}_n,\\
    \langle \sigma^-_n\rangle = 
	\sum_{a,b} \eta_p^a \eta_{p'}^b s^{(a,b)}_n,\quad 
    \langle \sigma^z_n\rangle =
	\sum_{a,b} \eta_p^a \eta_{p'}^b z^{(a,b)}_n,
    \end{gathered}\label{sm:eq:expansion}
\end{align}
where \(\alpha^{(0,0)}_n=s^{(0,0)}_n=0\), \(z_n^{(0,0)}=-1\) 
%for all \(n\) 
correspond to the normal state.
% Next we show how \Delta T is calculated in terms of these auxiliary variables...
% Also or before that: example of how to calculate dynmiacs of alpha's 

\subsection{Differential Transmission \texorpdfstring{$\Delta T_n$}{}}
We define the differential transmission as the 
normalized transmission \(T_n(\omega)\) of the probe pulse
%cavity transmission
with and without %the presence of 
the pump pulse~\cite{reitz2025nonlinear}:
\begin{align}
\Delta T_n (\omega)=T_n^{p-\mathrm{on}}(\omega)-T_n^{p-\mathrm{off}}(\omega).
\end{align}
%Here, \(T_n^{p-\mathrm{on/off}}\) is the linear transmission of the \textit{probe}
%when the pump pulse is on or off. In particular we will
%subtract from \(T_n^{p-\mathrm{on}}\) terms involving the pump only.
In particular, we omit from \(T_n^{p-\mathrm{on}}\) any terms involving the pump only
(e.g., \(\alpha^{(1,0)}_n\) below), to correctly obtain the probe transmission. 

\begin{figure}[t]
    \centering
    \includegraphics[width=0.95\linewidth]{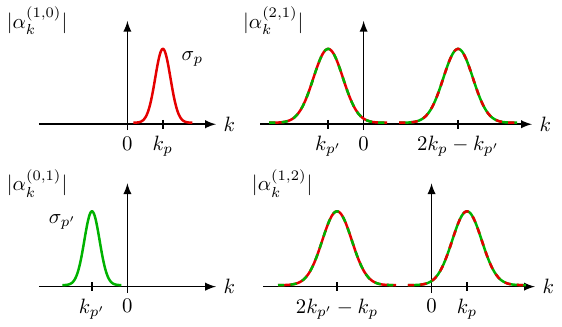}
    \caption{Momentum-space distributions of linear (left) and third-order cavity fields (right).
    The linear pump-induced cavity field $\alpha^{(1,0)}$ is centered around \(k_p\), while the linear probe-induced cavity field $\alpha^{(0,1)}$ is centered around \(k_{p'}\). At third order, the \((2,1)\) contribution contains the phase-matching channels
    \(k_p-k_p+k_{p'}=k_{p'}\) and \(k_p+k_p-k_{p'}\), where the first combination corresponds to the standard pump-probe signal,
    and the second to the so-called double-quantum-coherence signal, which is not of interest in this work.
    Since the detected signal is selected along the probe direction \(k_{p'}\), only the pump-probe contribution of \((2,1)\) contributes to the measured response. Note for the experiment considered in the Letter,
    \(k_p=-k_\pp=\pi/2\)~\(\mu\text{m}^{-1}\) and \(\abs{2k_p-k_\pp}=\abs{2k_\pp-k_p}=3k_p\).}
    \label{figuresm8}
\end{figure}

From the input-output relations for cavity transmission~\cite{steck2007},
the output field at each position \(n\)
is directly proportional to the internal cavity field \(\langle \tilde{a}_n^{\phantom{\dagger}}\rangle\). The 
transmitted intensities are proportional to the square of the output field
and so, keeping terms up to third order as well as only linear contributions in the probe field,
\begin{subequations}
\begin{align}
T_n^{p-\mathrm{off}}(\omega)&\propto (\kappa/2)^2|\alpha_n^{(0,1)}(\omega)|^2, \label{sm:eq:tpoff} \\
T_n^{p-\mathrm{on}}(\omega)&\propto (\kappa/2)^2|\alpha_n^{(0,1)}(\omega)+
\eta_p^2\alpha_n^{(2,1)} (\omega)|^2.\label{sm:eq:tpon} 
\end{align}
\end{subequations}
Since \(\eta_{p'} \ll \eta_p\), we neglect, in particular, the \((1,2)\) contribution, which is suppressed relative to the \((2,1)\) term by an additional power of the weak probe amplitude. Moreover, because the detected signal is selected along the probe propagation direction, higher-order contributions generally fulfill different phase-matching conditions and radiate into different momentum channels. They therefore do not contribute to the measured response considered here, as illustrated in Fig.~\ref{figuresm8}, which schematically shows the $k$-space distributions of the linear and third-order cavity fields as obtained from a Fourier space analysis of Eqs.~\eqref{eq:full-system}. 
We therefore obtain (keeping only the leading third-order correction to the differential transmission)
\begin{align}
\Delta T_n (\omega) \propto \eta_p^2 (\kappa^2/2)\mathrm{Re}[\alpha_n^{(0,1)}(\omega)\overline{\alpha}_n^{(2,1)}(\omega)],
\end{align}
which is Eq.~(10) of the Letter. Note we do not
normalize this quantity.

\section{Perturbative pump-probe equations}
As explained above, the leading contribution to  the nonlinear
differential transmission \(\Delta T_n\)  arises from terms second order in the
pump and first  order in the probe~\cite{reitz2025nonlinear}:  \(\alpha^{(0,1)}_n\) and
\(\alpha^{(2,1)}_n\). The  equations of motion for variables at a given order
depend on those of equal or lower order. Thus, in  principle, one requires the
dynamics of all  \(\alpha^{(a,b)}_n\), \(\sigma^{(a,b)}_n\), and
\(z^{(a,b)}_n\) for \(a \leq 2\), \(b \leq 1\).  

However, many of these variables remain trivial for  the initial conditions we
consider,  \(\langle \tilde{a}^{\phantom{\dagger}}_n \rangle =  \langle
\sigma^-_n \rangle = 0\),  \(\langle \sigma^z_n \rangle = -1\) at \(t = 0\).
This ensures all \(\alpha^{(a,b)}_n\),  \(\sigma^{(a,b)}_n\), and
\(z^{(a,b)}_n\) are initially  zero, except for \(z^{(0,0)}_n \equiv -1\).  The
resulting nontrivial set of equations is:  
\begin{widetext}
\begin{equation}
\begin{aligned}
		\partial_t \alpha_n^{(1,0)} &=
		-\left( i {\omega}^{\vphantom{\dagger}}_{c} +
		\kappa/2 - iC \partial^2_n \right)\alpha_n^{(1,0)}
		- i \Omega \sigma_n^{(1,0)} + D^p_n f_p(t)
		e^{-i {\omega}^{\vphantom{\dagger}}_{p} t}, \\
		\partial_t \sigma_n^{(1,0)} &=
		-( i {\omega}^{\vphantom{\dagger}}_{0} + \gamma_\phi/2)\sigma_n^{(1,0)}
		+ i \Omega  z_n^{(0,0)} \alpha_n^{(1,0)}, \\
		\partial_t z_n^{(2,0)} &= 
		- 4 \Omega \Im \left[\sigma_n^{(1,0)}  \overline{\alpha}_n^{(1,0)}\right] ,\\
		\partial_t \alpha_n^{(0,1)} &=
		-\left( i {\omega}^{\vphantom{\dagger}}_{c} + \kappa/2 -
		iC \partial^2_n \right)\alpha_n^{(0,1)}
		- i \Omega \sigma_n^{(0,1)} + D^\pp_n f_\pp(t)
		e^{-i {\omega}^{\vphantom{\dagger}}_{\pp} t}, \\
		\partial_t \sigma_n^{(0,1)} &=
		-( i {\omega}^{\vphantom{\dagger}}_{0} + \gamma_\phi/2)\sigma_n^{(0,1)}
		+ i \Omega  z_n^{(0,0)} \alpha_n^{(0,1)}, \\
		\partial_t z_n^{(1,1)} &=
		- 4 \Omega \Im \left[ 
		\sigma_n^{(1,0)}\overline{\alpha}_n^{(0,1)}+
		\sigma_n^{(0,1)}\overline{\alpha}_n^{(1,0)}
		\right], \\
        \partial_t \alpha_n^{(2,1)} &=
		-\left( i {\omega}^{\vphantom{\dagger}}_{c} + \kappa/2 -
		iC \partial^2_n \right)\alpha_n^{(2,1)}
		- i \Omega \sigma_n^{(2,1)}, \\
		\partial_t \sigma_n^{(2,1)} &=
		-( i {\omega}^{\vphantom{\dagger}}_{0} + \gamma_\phi/2)\sigma_n^{(2,1)}
		+ i \Omega \left( 
			z_n^{(0,0)}\alpha_n^{(2,1)} 
			+ z_n^{(1,1)}\alpha_n^{(1,0)} 
			+ z_n^{(2,0)}\alpha_n^{(0,1)} 
		\right).
\end{aligned}
\label{eq:full-system}
\end{equation}
\end{widetext}
Here, we have explicitly included the pump and probe  
driving terms from \cref{sm:eq:ain}. As in the Letter, an overbar
is used to denote a complex conjugate. These equations were  
obtained by first deriving the Heisenberg equations of motion  
for the single-operator expectations  
\(\langle \tilde{a}^{\phantom{\dagger}}_n \rangle\),  
\(\langle \sigma^-_n \rangle\), and  
\(\langle \sigma^z_n \rangle\) from the master equation  
\eqref{am:eq:ME}. The mean-field approximation,  
\(
    \langle \tilde{a}^{\phantom{\dagger}}_n \sigma^z_n \rangle
    \approx \langle \tilde{a}^{\phantom{\dagger}}_n
    \rangle \langle \sigma^z_n \rangle, \
    \langle \tilde{a}^{\phantom{\dagger}}_n \sigma^+_n \rangle
    \approx \langle \tilde{a}^{\phantom{\dagger}}_n
    \rangle \langle \sigma^-_n \rangle^*,
\)
was applied before making the expansion described by
\cref{sm:eq:expansion} and  
collecting terms of identical order in \(\eta_p\) and \(\eta_{\pp}\).

\section{Transport Velocity from Counter-Propagating Pulses}

\begin{figure}
    \centering
    \includegraphics[width=\linewidth]{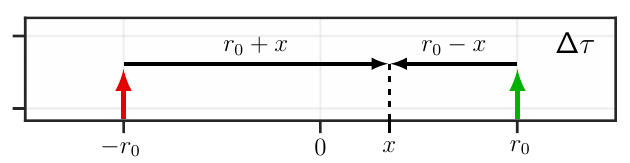}%
    \caption{ Calculating transport velocities in the counter-propagating experiment. Red and green arrows mark the initial spatial locations of the pump and probe spots, respectively, which
    are initially separated a distance
    \(2r_0=100\ \mu\text{m}\).}
    \label{fig:sm1}
\end{figure}

In this section, we show how polariton transport velocities are extracted from
the  motion  of the differential signal \(\Delta T_n \)  in Fig.~3 of the Letter
%as a function of \(\Delta \tau\)
arising from counter-propagating pump and probe pulses.

Suppose the probe is delayed by \(\Delta \tau\) and that after a time \(t\) the
pump has advanced to a position \(x\) to the right of the origin
(see sketch in \cref{fig:sm1}).  The total distance traversed by the pump is \(r_0+x\)
over the time interval \(t\), giving a transport velocity of
\(v_{k_p}^{\phantom{\dagger}}= (r_0+x)/t\)>0.  The probe, on the other hand,
traveled a distance \(r_0-x\) in a time \(t-\Delta \tau\) reduced by the delay,
giving \(v_{k_{p'}}^{\phantom{\dagger}}= -(r_0-x)/(t-\Delta \tau)<0\). Here, we used the symmetry of the polariton dispersion about \(k=0\), which implies that the
group velocities have equal magnitude and opposite sign for \(k_{p'}=-k_p\), i.e.,
\(
v_{k_{p'}}=-v_{k_p}
\).

From the probe equation,
\begin{align}
 v_{k_p}^{\phantom{\dagger}}  
(t-\Delta \tau)  = r_0-x. 
\end{align}
Using the pump equation
\(v_{k_p}^{\phantom{\dagger}} t =r_0+x\), we obtain:
\begin{align} (r_0+x) -
v_{k_p}^{\phantom{\dagger}} \Delta \tau &= r_0-x \\\nonumber \Rightarrow
v_{k_p}^{\phantom{\dagger}} &= \frac{2x}{\Delta \tau}. 
\end{align}
Thus, the transport velocity is determined as \textit{twice} the speed
\(dx/d(\Delta \tau)\) of the signal arising from the counter-propagating pulses.

To connect this result to the velocities \(v_{k_p}^{\text{peak}}\) and \(v_{k_p}^{\text{rms}}\)
reported in Fig.~4 of the Letter, note that the differential transmission
signal is an extended object, and therefore a choice must be made to 
define \(x\) during the experiment. Choosing the maximum absolute value of
the differential signal gives the peak velocity  \(v_{k_p}^{\text{peak}}\),
whilst choosing the root-mean-square displacement of the signal yields
\(v_{k_p}^{\text{rms}}\).

\section{Transport with Exciton Hopping and Exciton-Exciton
Annihilation}

%In this section we show the effect of exciton hopping
%and pair annihilation on the polariton transport in the model.

In this section, we provide results for the polariton transport in the presence of additional processes that may become important in realistic materials such as organic semiconductors: exciton hopping and exciton–exciton pair annihilation \cite{kohler2015electronic, mikhnenko2015exciton}. Exciton hopping enables energy transfer between neighboring sites, while exciton–exciton annihilation introduces nonlinear dissipation. \\

\subsection{Derivation of Perturbative Equations}
These processes may be readily included in our framework
with the extended model (cf.~\cref{sm:eq:ME2})
\begin{align}
\begin{split}
		\partial_t \rho = &-i \bigl[ H+ H_{\text{hop}}  , \rho \bigr]
	+ \sum_{k=1}^{N_k} \kappa \mathcal{L}[a^{\vphantom{\dagger}}_{k}] \\
    &+ \sum_{n=1}^{N_k} \sum_{\x=1}^{N_s} \frac{\gamma_\phi}{4}\mathcal{L}[\sigma^z_{n\x}]\\
    &+ \sum_{n=1}^{N_k} \sum_{\x,\y=1}^{N_s} \gamma^{\text{ee}}\mathcal{L}[\sigma^-_{n\x}\sigma^-_{n\y}].
	\end{split}\label{sm:eq:ME2}
\end{align}
The parameter \(\gamma^{\text{ee}}\) sets the rate of exciton-exciton
annihilation, an incoherent process which becomes relevant for many organic
materials at high optical intensities under intense pulse excitation \cite{akselrod2010exciton}, while 
\( H_{\text{hop}}\) describes coherent hopping between neighboring coarse-grained sites,
\begin{align}
 H_{\text{hop}}  = - J\sum_{n=1}^{N_k} \sum_{\x,\y=1}^{N_s} 
\bigl( 
\sigma^+_{n\x} \sigma^-_{n+1,\y} + \text{H.c}
\bigr),\label{sm:eq:Hhop}
\end{align}
with hopping amplitude \(J\). The importance of inter-site hopping is set by whether the exciton diffusion length is comparable to the coarse-graining scale $\Delta r$.
%[Argue in what cases this is relevant or not, e.g. small length scales, 
%long times ?]
% J is t_hop in the code, want to avoid using t or tau 

Since the additional terms involve only molecular operators, the equation
of motion for the photonic variable, Eq.~(4a) of the main text, remains unchanged. 
Those for the spin variables acquire additional terms proportional to \(J\) and
\(\gamma^{\text{ee}}\). Using the same coarse-graining described
above, under the mean-field approximation, one has
\begin{subequations}
\begin{align}
\begin{split}
		\partial_t \langle \sigma^-_{n} \rangle  &= -
		\bigl( i \omega_0^{\vphantom{\dagger}} +
\gamma_\phi/2 \bigr) \langle \sigma^-_{n} \rangle
+ i \Omega
\langle \sigma^z_{n} \rangle \langle \tilde{a}^{\vphantom{\dagger}}_{n} \rangle
\\
&\phantom{=}\ -i J N_s \langle \sigma^z_n \rangle \left(
\langle \sigma_{n+1}^- \rangle +
\langle \sigma_{n-1}^- \rangle 
\right)
\\
&\phantom{=}\
- (\gamma^{\text{ee}}/2) (N_s-1) \langle \sigma_n^- \rangle 
\left( 1 + \langle \sigma^z_n \rangle \right),
\end{split}
		\label{sm:eq:mfxb}	
		\\
\begin{split}
		\partial_t \langle \sigma^z_{n} \rangle &= 
		- 4 \Omega
		\Im \left[ \langle \sigma^-_{n} \rangle\langle
				\tilde{a}^{\vphantom{\dagger}}_{n}\rangle^*
		\right]
\\
&\phantom{=}\
-4 J N_s \Im\left[ 
\langle \sigma_n^- \rangle^\ast \left(\langle \sigma_{n+1}^- \rangle +
\langle \sigma_{n-1}^- \rangle  \right)
\right]
\\
&\phantom{=}\
- \gamma^{\text{ee}} (N_s-1) \left( 1 + 2 \langle \sigma^z_n \rangle + \langle
  \sigma^z_n \rangle^2 \right)
. 
\end{split}
		\label{sm:eq:mfxc}	
\end{align}
\end{subequations}

It remains to determine revised equations for the coefficients 
\(\alpha_n^{(a,b)}\), \(\sigma^{(a,b)}\), \(z^{(a,b)}_n\) in the 
perturbation expansion. Those for the photonic variable are unchanged
from \cref{eq:full-system}, whilst those for the spin variables acquire
new local and non-local terms. Noting \(z_n^{(0,0)}\equiv -1\) still,
the resulting modified equations read
\begin{widetext}
\begin{equation}
\begin{aligned}
		\partial_t \sigma_n^{(1,0)} &=
		-( i {\omega}^{\vphantom{\dagger}}_{0} + \gamma_\phi/2)\sigma_n^{(1,0)}
		- i \Omega  \alpha_n^{(1,0)}
        +iJ N_s  \left(\sigma^{(1,0)}_{n-1} + \sigma^{(1,0)}_{n+1} \right)
        , \\
		\partial_t z_n^{(2,0)} &= 
		- 4 \Omega \Im \left[\sigma_n^{(1,0)}  \overline{\alpha}_n^{(1,0)}\right]
        -4JN_s \Im \left[\overline{\sigma}_n^{(1,0)}
        \left(\sigma^{(1,0)}_{n-1} + \sigma^{(1,0)}_{n+1} \right)
        \right]
        -2 \gamma^{\text{ee}}(N_s-1)
        z^{(2,0)}_n
        ,\\
		\partial_t \sigma_n^{(0,1)} &=
		-( i {\omega}^{\vphantom{\dagger}}_{0} + \gamma_\phi/2)\sigma_n^{(0,1)}
		- i \Omega  \alpha_n^{(0,1)}
         +iJ N_s  \left(\sigma^{(0,1)}_{n-1} + \sigma^{(0,1)}_{n+1} \right)
        , \\
		\partial_t z_n^{(1,1)} &=
		- 4 \Omega \Im \left[ 
		\sigma_n^{(1,0)}\overline{\alpha}_n^{(0,1)}+
		\sigma_n^{(0,1)}\overline{\alpha}_n^{(1,0)}
		\right] \\
        &\phantom{=}\
         -4JN_s \Im \left[\overline{\sigma}_n^{(1,0)}
        \left(\sigma^{(0,1)}_{n-1} + \sigma^{(0,1)}_{n+1} \right)+
        \overline{\sigma}_n^{(0,1)}
        \left(\sigma^{(1,0)}_{n-1} + \sigma^{(1,0)}_{n+1} \right)
        \right] \\
        &\phantom{=}\ -2 \gamma^{\text{ee}} (N_s-1) z_n^{(1,1)}
        , \\
		\partial_t \sigma_n^{(2,1)} &=
		-( i {\omega}^{\vphantom{\dagger}}_{0} + \gamma_\phi/2)\sigma_n^{(2,1)}
		+ i \Omega \left( 
			 z_n^{(1,1)}\alpha_n^{(1,0)} +
			 z_n^{(2,0)}\alpha_n^{(0,1)}
			-\alpha_n^{(2,1)} 
		\right) \\
        &\phantom{=}\ 
        + iJN_s\left[ \left(\sigma^{(2,1)}_{n-1} + \sigma^{(2,1)}_{n+1} \right)
       - z^{(2,0)}_n \left(\sigma^{(0,1)}_{n-1} + \sigma^{(0,1)}_{n+1} \right)
       -  z^{(1,1)}_n\left(\sigma^{(1,0)}_{n-1} + \sigma^{(1,0)}_{n+1} \right)
       \right]\\
       &\phantom{=}\ 
       - (\gamma^{\text{ee}}/2)(N_s-1)
       \left(
       z_n^{(1,1)} s^{(1,0)}_n
       +
       z_n^{(2,0)} s^{(0,1)}_n
       \right).
\end{aligned}
\label{eq:full-system-hop-eea}
\end{equation}
\end{widetext}
Hopping processes introduce non-local terms in the spin variables, enabling transport between neighboring sites at an effective rate \(JN_s\) proportional to the number of emitters per site \(N_s\). However, note that the
site spacing \(\Delta r = L/N_k \sim 0.3\ \mu\)m is such that this transport may not be
discernible on the scale of the polariton-assisted transport, 
even when \(J N_s\) is comparable to the collective light-matter coupling.
On the other hand, exciton-exciton annihilation processes introduce additional loss terms for second-order and higher-order spin variables, with rates that scale similarly with \(N_s\).

\subsection{Pump-only Dynamics With Hopping}

\begin{figure}
    \centering
    \includegraphics[width=\linewidth]{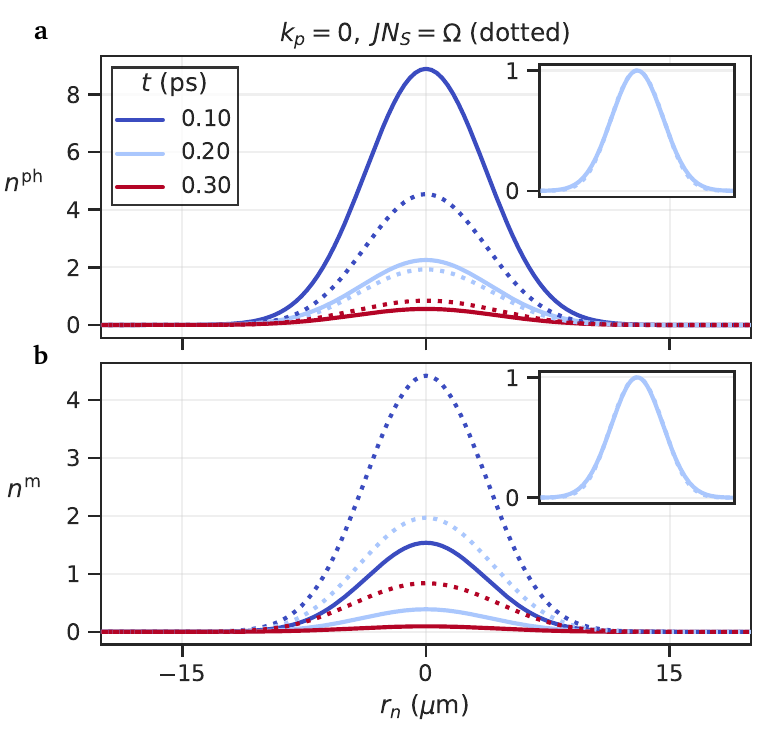}%
    \caption{%
    Dynamics of photonic 
    \(n^{\text{ph}}\) and molecular \(n^{\text{m}}\)
    populations  following a
    Gaussian pump pulse at \(r_n=0\) with
    central wavevector \(k_p=0\).
    Solid lines: without excitonic
    hopping \(J=0\); dotted lines: with hopping
    \(JN_s=\Omega\). System parameters
    (except \(J\)) match those in Fig.~2
    of the Letter.
    % Note that for single-pulse dynamics, 
    % the effect of exciton-exciton annihilation 
    % not shown in negligible for \(\gamma_{\text{ee}} N_s \sim \kappa\)
    %(not shown) is negligible under weak excitation.
    Insets include normalized copies of the \(t=0.2\)~ps curves. 
    This shows, in particular, that while the magnitude of \(n^m\)
    changes significantly in the presence of hopping, its spatial distribution
    does not.
    % there is just a scale factor between the solid and dotted lines
    }
    \label{fig:sm2}
\end{figure}

\Cref{fig:sm2} shows dynamics following a single Gaussian pump pulse with
central wavevector \(k_p=0\) in the absence (solid line) and presence (dashed
line) of excitonic hopping. We take \(J\) such that the effective intersite
coupling \(J N_s\) equals the collective light-matter coupling
\(\Omega=g\sqrt{N}\). 
% could reference 0-50mev typical for organic materials

The most noticeable effect of hopping here is a change in the composition of the
light-matter excitation: the introduced excitonic dispersion brings excitonic
states near \(k_p=0\) down in energy towards the cavity dispersion,
and the polariton becomes more excitonic.  Consequently, when \(J>0\),
at short times
the molecular populations \(n^{\text{m}}\) (bottom row of \cref{fig:sm2}) are
larger and the photonic populations \(n^{\text{ph}}\) (top row)
are smaller. At
late times, the photon populations in the presence of hopping
may also be larger. This is because there is no
molecular loss (\(\gamma_{\text{ee}}=0\)), so making the polariton more
excitonic effectively increases its lifetime by protecting against cavity loss
\(\kappa\). 

Given that the pump has central wavevector \(k_p=0\), there is no net
movement of the excitation, whilst outward movement at the tails
due to polaritons with wavevectors \(k\neq 0\) is not significant, even with
\(J>0\): the characteristic length scale for hopping of \(\Delta r \sim 0.3\
\mu\)m is sufficiently small that the spatial spread 
remains essentially unchanged over the polariton lifetime (see inset).

With exciton-exciton annihilation \(\gamma_{\text{ee}}N_s \sim \kappa\),
the effect on single-pulse dynamics (not shown in \cref{fig:sm2}) is
negligible. However, it can still have a dramatic effect on the dynamics
of higher-order signals, as we show below.

% \begin{itemize}
%     \item Could comment on typical hopping values 
%     (but uncertain, easier to leave out)
%     \item Not like-for-like as \(J\) changes eigenstates
%     (redshift, see figures below), more excitonic
%     \item change in spread of wavepacket not noticeable (see inset)
%     \item For single-pulse dynamics, effect of 
%     EEA is negligible in weak excitation limit (right?)
% \end{itemize}

\subsection{Pump-probe Experiment with Hopping and Exciton-Exciton Annihilation}

\begin{figure}
    \centering
    \includegraphics[width=\linewidth]{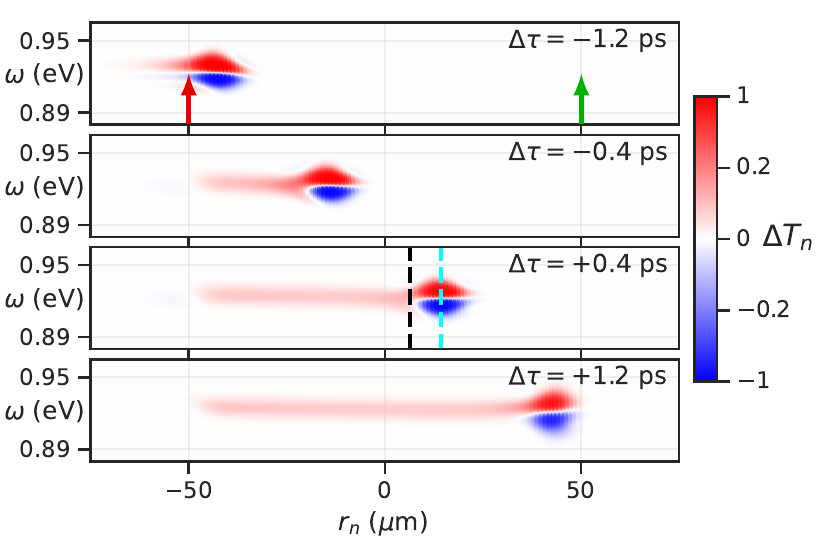}%
    \caption{Differential transmission \(\Delta T_n\) 
    following evolution of a LP
	excitation
    in the pump-probe experiment. Copy of Fig.~3 of the main text.}
    \label{fig:sm_repeat_3}
\end{figure}
\begin{figure}
    \centering
    \includegraphics[width=\linewidth]{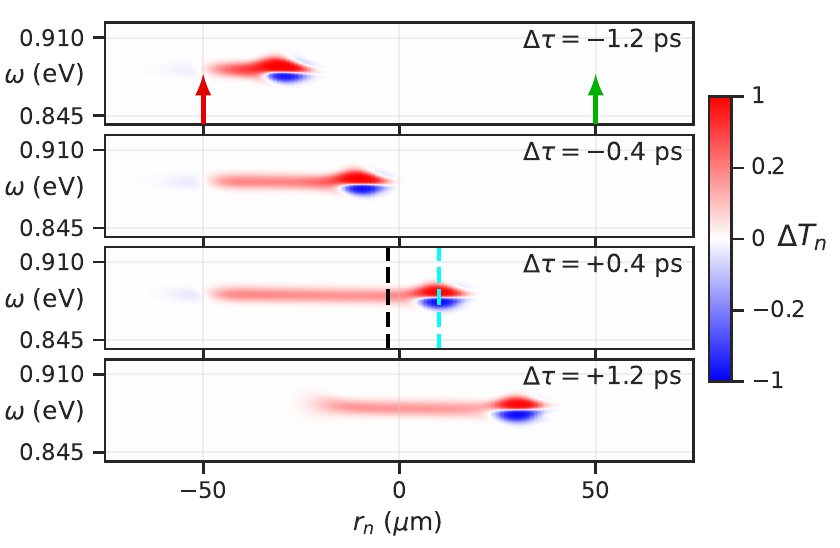}%
    \caption{Differential transmission \(\Delta T_n\) following evolution of a LP
	excitation in the presence of exciton hopping \(J N_s=\Omega\).
    All other parameters and the experimental setup matches that used for 
    Fig.~3 of the Letter. As in Fig.~3, black and cyan dashed lines
    mark the rms position and center (maximum) of the excitation (as
    the delay \(\Delta \tau\) at which the signal is revealed at
    the pump spot at \(r_n=-50\ \mu\)m changes, one can not directly
    compare these lines between figures). Note the energy scale is red-shifted 
    compared to \cref{fig:sm_repeat_3,fig:sm4}.}
    \label{fig:sm3}
\end{figure}
\begin{figure}
    \centering
    \includegraphics[width=\linewidth]{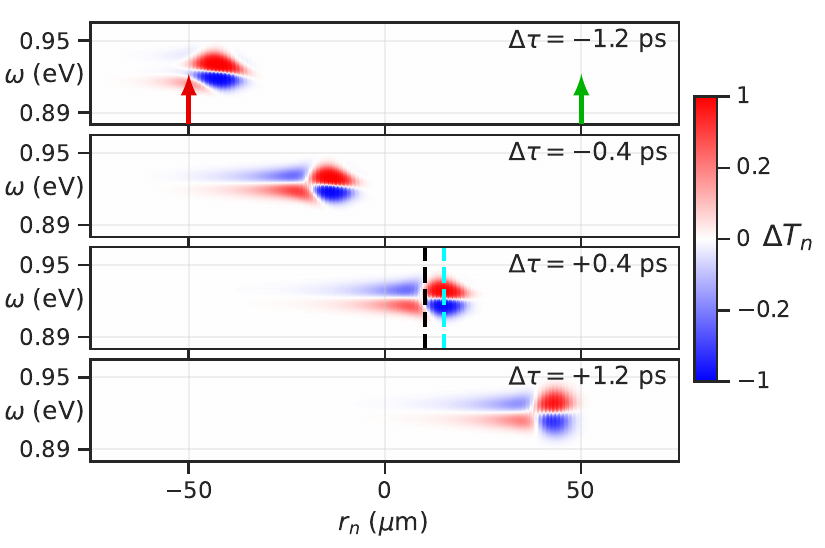}%
    \caption{Differential transmission \(\Delta T_n\) following evolution of a LP
	excitation in the presence of exciton-exciton annihilation
       \(\gamma_{\text{ee}} (N_s-1)=\kappa/8\).
       All other parameters and the experimental setup matches that used for 
    Fig.~3 of the Letter. As in Fig.~3, black and cyan dashed lines
    mark the rms position and center (maximum) of the excitation.
       }
    \label{fig:sm4}
\end{figure}

We now present results for the pump-probe experiment described in the Letter
including hopping \(J\) or exciton-exciton annihilation
\(\gamma_{\text{ee}}\). For comparison, we repeat Fig.~3 from the Letter
(\(J=0=\gamma_{\text{ee}}=0\)) in \cref{fig:sm_repeat_3}. 

\Cref{fig:sm4} shows two
main effects of including hopping \(J N_s=\Omega\).
First, the signal is redshifted since
the excitonic dispersion is brought down towards the cavity dispersion at
\(k=k_p=\pi/2\ \mu\text{m}^{-1}\) 
(\(\Delta E_k = -2JN_s \cos(k \Delta r) |Y_k|^2\) in
first-order perturbation theory,
where \(Y_k\) is the matter 
Hopfield coefficient).
 Second, the transport slows:
at \(\Delta \tau=-1.2\)~ps the probe pulse finds the excitation to the right of
the initial pump position \(r_n=-50\ \mu\)m, and at \(\Delta \tau=1.2\)~ps the
signal has still not reached the probe position \(r_n=50\ \mu\)m (compare \cref{fig:sm_repeat_3}).
We note that, at the last time, \(\Delta \tau= 1.2\)~ps, the tail of the signal
appears to disappear. This is likely due the higher excitonic content reducing
the probe pulse's ability to detect the full spatial extent of the signal.

\begin{figure}
    \centering
    \includegraphics[width=\linewidth]{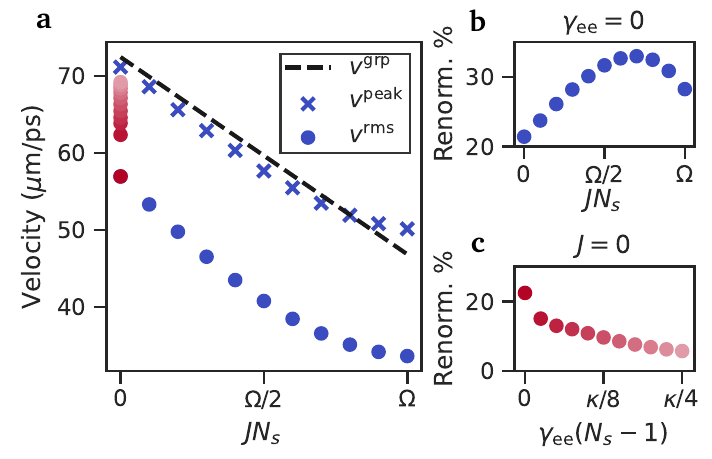}%
    \phantomsubfloat{\label{fig:sm5a}}%
    \phantomsubfloat{\label{fig:sm5b}}%
    \phantomsubfloat{\label{fig:sm5c}}%
	\vspace{-2\baselineskip}%
    \caption{%   \textit{Transport renormalisation.}
	(a) Blue: velocity of the circular feature (\(v^{\text{peak}}\),
	crosses) and rms displacement (\(v^{\text{rms}}\), circles) for 
    various hopping strengths \(JN_s\) at fixed pump wavevector
    \(k_p=\pi/2\ \mu\text{m}^{-1}\) and \(\gamma_{\text{ee}}=0\).
    The velocities were calculated via the same procedure used for
    Fig.~4 in the Letter.
    The dashed curve shows the theoretical group velocity
	\(\tilde{v}^{\text{grp}}\) at \(k_p\), modified according
    to the perturbation \(JN_s\).
    (b) Increasing renormalization 
     \(1-v^{\text{rms}}/\tilde{v}^{\text{grp}}\) is observed with 
     increasing \(JN_s\) until \(JN_s \approx 2\Omega/3\), beyond
     which the probe pulse begins to fail to reach the pump spot
     and perturbation theory breaks down.
	(a) Red: \(v^{\text{rms}}\) at 
    \(k_p=\pi/2\ \mu\text{m}^{-1}\) and \(J=0\) for 
    \(\gamma_\text{ee}\) increasing from \(0\) to \(\kappa/4\)
    (upwards, lighter red).
	The corresponding renormalization trend is shown in (c).
    }
    \label{fig:sm5}
\end{figure}

Two main factors contribute to the modified transport behavior. One mechanism is
that hopping increases the excitonic character of the polariton, making
transport more diffusive.  A second is that the modified dispersion relation
changes the group velocity.  To distinguish these factors, we calculate the
ratio \(v^{\text{obs}}/\tilde{v}^{\text{grp}}\) relative to the group velocity
\(\tilde{v}^{\text{grp}}\) of the dispersion modified by the hopping according
to first-order perturbation theory at \(k_p=\pi/2\) \(\mu\text{m}^{-1}\).  The
results, presented in \crefrange{fig:sm5a}{fig:sm5b} for \(0\leq J N_s \leq
\Omega\), show that, while increasing \(J\) decreases the group velocity
\(\tilde{v}^{\text{grp}}\) at this \(k_p\), the transport velocity
\(v^{\text{obs}}\) decreases more than can be accounted for by the decrease in
\(\tilde{v}^{\text{grp}}\) alone, i.e., the renormalization increases.  This
trend can be explained by the increasing exciton content (i.e., changing
Hopfield coefficients) of the propagating excitation with increasing hopping.
The subsequent decrease at large \(JN_s\gtrsim 2\Omega/3 \) in \cref{fig:sm5b}
is likely due to the probe pulse failing to cover the full extent of the signal
(\(\Delta \tau= 1.2\)~ps in \cref{fig:sm4}), as well as by the breakdown of the
perturbation theory when \(JN_s\) is no longer small.

Finally, the addition of exciton-exciton annihilation has a surprising effect on
transport: by providing a loss mechanism for dark, dephased molecular excitations, the
ballistic proportion of the propagating excitation increases,
and hence does the
effective (i.e., rms)
transport velocity. The presence of this effect, i.e., the 
depletion of
the `dragged' molecular tail, is highly sensitive to the value of
\(\gamma_{\text{ee}}\). In \cref{fig:sm5c}, we see a sudden decrease
in renormalization as \(\gamma_{\text{ee}}\ll \kappa\) is switched on from zero,
then a steady decrease as \(\gamma_{\text{ee}} \to \kappa\).
This further demonstrates the complex range of
behaviors in organic materials, where observed transport is not simply determined
by group velocities.  Instead, one must also consider the spectroscopic
observable being measured, the nature of the excitation, and relevant loss
processes. We believe the effect of exciton-exciton annihilation is an
under-explored aspect in models of organic polariton transport, as a process
that is often unavoidable in real experiments \cite{akselrod2010exciton}.

\section{Molecular Disorder}%
In this section, we show how static molecular disorder can be included in the model,
allowing for  more realistic descriptions of microcavity experiments
involving disordered organic materials.%

To incorporate molecular disorder, we consider the generalized Tavis-Cummings
Hamiltonian,
\begin{align}
\begin{split}
	H_{} &= \sum_{k=1}^{N_k}
	{\omega}^{\vphantom{\dagger}}_{k} a^{\dagger}_{k}a^{\vphantom{\dagger}}_{k}
    +  \sum_{n=1}^{N} \frac{\omega_n^{\vphantom{\dagger}}}{2} \sigma^z_{n}\\ &+   \sum_{k=1}^{N_k}\sum_{n=1}^{N}
	g_n(a^{\vphantom{\dagger}}_{k}\sigma^+_{n}e^{ikr_n} + 
    \text{H.c.}),
\end{split}
\label{sm:eq:Hd}
\end{align}
in which the energy \({\omega}^{\vphantom{\dagger}}_{0}\to {\omega}^{\vphantom{\dagger}}_{n}\) and
coupling \(g\to g_n\) became emitter-dependent (note here \(n=1,\ldots, N\)
enumerates the entire molecular ensemble, not spatial position).

To coarse-grain the one-dimensional molecular system, we now
assign three indices to the spin variables: \(n\), for the 
site position \(r_n\) (\(n=1,\ldots, N_k\)); \(i\), for the
class of molecule with energy and coupling 
\({\omega}^{\vphantom{\dagger}}_{i}, g_i\)  (\(i=1,\ldots, M\));
and \(\x_i\), for individual molecules within
class \(i\). Here \(\x_i=1,\ldots, \mu_i N_s\) where \(N_s\) is the number of molecules
per site (as before) and \(\mu_i\) is the fraction of molecules with energies
and couplings within a range about \( {\omega}^{\vphantom{\dagger}}_{i}\) and \(g_i\).
The Hamiltonian is then expressed as (cf. \cref{sm:eq:H2}):
\begin{align}
	\begin{split}
	H_{} &= \sum_{k=1}^{N_k}
	{\omega}^{\vphantom{\dagger}}_{k} a^{\dagger}_{k}a^{\vphantom{\dagger}}_{k}
	+
\sum_{n=1}^{N_k}\sum_{i=1}^{M} \sum_{\x_i=1}^{\mu_iN_s} 
	\frac{\omega_i}{2} \sigma^z_{ni\x_i} \\
    &+  \sum_{k=1}^{N_k}\sum_{n=1}^{N_k}\sum_{i=1}^{M} \sum_{\x_i=1}^{\mu_iN_s} 
	g_i(a^{\vphantom{\dagger}}_{k}\sigma^+_{ni\x_i}e^{ikr_n} + 
    \text{H.c.})
	,\end{split}
\label{sm:eq:H2d}
\end{align}
and the master equation,
\begin{align}
		\begin{split}
		\partial_t \rho = &-i \bigl[ H , \rho \bigr]
	+ \sum_{k=1}^{N_k} \kappa \mathcal{L}[a^{\vphantom{\dagger}}_{k}] \\
	&+ \sum_{n=1}^{N_k}\sum_{i=1}^{M} \sum_{\x_i=1}^{\mu_iN_s} \frac{\gamma_\phi}{4}\mathcal{L}[\sigma^z_{ni\x_i}].
\end{split}
	\label{sm:eq:ME2bd}
\end{align}
In the following, we set \(g_i=g\) to obtain results for the case of energetic disorder only, but 
results with inhomogeneous couplings can be derived in the same way.

The derivation of the mean-field equations is identical to before up to evaluation
of sums over the molecular indices \(\x_i\):
\begin{subequations}
\begin{align}
\begin{split}
    \partial_t \langle a^{\phantom{\dagger}}_k \rangle
    &=  	-\bigl( i {\omega}^{\vphantom{\dagger}}_{k} + \kappa/2 \bigr) \langle
		a^{\vphantom{\dagger}}_{k} \rangle\\
&- i g \sum_{n=1}^{N_k}\sum_{i=1}^{M} \sum_{\x_i=1}^{\mu_i N_s} e^{-ikr_n} \langle \sigma^-_{ni\x_i} \rangle ,
\end{split} \label{sm:eq:mf2aD} \\
\begin{split}
\partial_t \langle \sigma_{ni\x_i}^-\rangle
&= -\bigl( i \omega^{\phantom{\dagger}}_i +  \gamma_\phi/2 \bigr) \langle \sigma_{ni\x_i}^- \rangle \\
&+i g \langle \sigma^z_{ni\x_i} \rangle \sum_{k=1}^{N_k}
e^{ikr_n}  \langle   a^{\phantom{\dagger}}_k \rangle,
\end{split} \label{sm:eq:mf2bD} \\
\partial_t \langle \sigma_{ni\x_i}^z\rangle
&=  -4g\sum_{k=1}^{N_k}\Im[\langle \sigma_{ni\x_i}^-  \rangle \langle a^{\phantom{\dagger}}_k\rangle^*
 e^{-ikr_n}]  .
\label{sm:eq:mf2cD}
\end{align}
\end{subequations}

It is no longer the case that all molecules on a given site have the same properties. However,
one can still sum over all molecules of the same \textit{class} within a site in \cref{sm:eq:mf2aD}:
\begin{subequations}
\begin{align}
\begin{split}
    \partial_t \langle a^{\phantom{\dagger}}_k \rangle
    &=  	-\bigl( i {\omega}^{\vphantom{\dagger}}_{k} + \kappa/2 \bigr) \langle
		a^{\vphantom{\dagger}}_{k} \rangle\\
&- i g N_s\sum_{n=1}^{N_k} e^{-ikr_n} \sum_{i=1}^M \mu_i \langle \sigma^-_{ni} \rangle ,
\end{split} \label{sm:eq:mf3a}
\end{align}
\end{subequations}
where \(\langle \sigma_{ni}^\alpha \rangle\) denotes
the expectation for any one of the 
spin variables at \(r_n\) with energy \({\omega}^{\vphantom{\dagger}}_{i}\).
After transforming the photon variables to real space and rescaling
(\(\tilde{a}_n^{\phantom{\dagger}} = \sqrt{N_k/N}a^{\phantom{\dagger}}_n \)),
the final equations are
\begin{subequations}
\label{eq:mfD}
\begin{align}
\begin{split}
    		\partial_t \langle \tilde{a}^{\vphantom{\dagger}}_{n}\rangle  &=
- \bigl(i {\omega}^{\vphantom{\dagger}}_{c}+ \kappa/2\bigr)
\langle \tilde{a}^{\vphantom{\dagger}}_n \rangle  
- i \Omega \sum_{\smash{i=1}}^M \mu_i  \langle \sigma^-_{ni} \rangle  \\
&\phantom{=}\
+iC \partial_n^2 \langle \tilde{a}^{\vphantom{\dagger}}_{n} \rangle
+ \sqrt{\kappa} \langle \tilde{a}_n^{\text{in}} \rangle,
\end{split}\label{eq:mfaD}\\
		\partial_t \langle \sigma^-_{ni} \rangle  &= -
		\bigl( i \omega_i^{\vphantom{\dagger}} +
\gamma_\phi/2 \bigr) \langle \sigma^-_{ni} \rangle
+ i \Omega
\langle \sigma^z_{ni} \rangle \langle \tilde{a}^{\vphantom{\dagger}}_{n} \rangle ,
		\label{eq:mfbD}	
		\\
		\partial_t \langle \sigma^z_{ni} \rangle &= 
		- 4 \Omega
		\Im \left[ \langle \sigma^-_{ni} \rangle\langle
				\tilde{a}^{\vphantom{\dagger}}_{n}\rangle^*
		\right],\label{eq:mfcD}
\end{align}
\end{subequations}
with \(\Omega=g\sqrt{N}\) and \(C=c^2/2 \omega^{\phantom{\dagger}}_c\).

Therefore, the number of mean-field equations scales with the number of classes one considers
when grouping the molecular energies into bins. Still, the low complexity of \crefrange{eq:mfaD}{eq:mfcD}
makes is possible to calculate dynamics for tens or even hundreds
of bins and hence achieve a fine resolution of molecular energy. 

To illustrate the above, we consider \(M=11\) bins sampling a Gaussian distribution of molecular energies centered around \(\omega_0=1.0\)~eV with standard deviation \(\sigma_m\). Introducing
energetic disorder modifies the polariton group velocities in a non-trivial way. To mitigate this effect
and so isolate any disorder-induced renormalization of the transport properties, 
we select a slightly lower pump and probe wavevector than that used
in the Letter, \(k_{p}=-k_{p'}=2 \pi/5\)~\(\mu\text{m}^{-1}\). At this wavevector,
the theoretical lower polariton group velocity \(\partial_k\omega^{\text{LP}}(k_p;\omega_m)\) is more
symmetrical about changes in molecular energy \(\omega_0\), thereby reducing artifacts arising from non-symmetric effects.

\begin{figure}
    \centering
    \includegraphics[width=\linewidth]{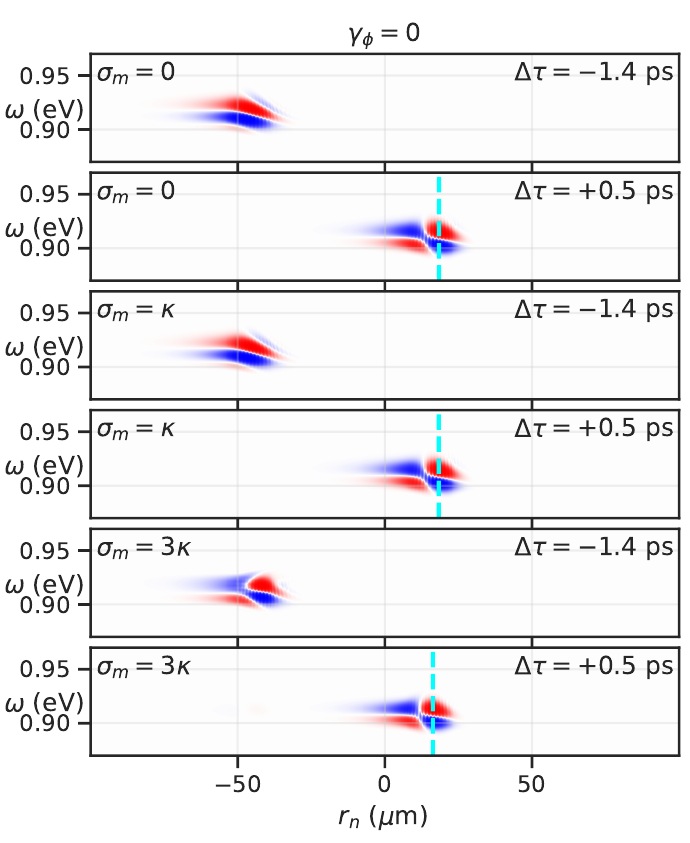}%
    \caption{%
    Differential transmission \(\Delta T_n\) showing position of 
    the LP excitation in the pump-probe experiment at two successive times
    with \(k_{p}=-k_{p'}=2 \pi/5\)~\(\mu\text{m}^{-1}\) and \(\gamma_\phi=0\),
    for three different disorder widths \(\sigma_m\). Dashed cyan lines
    indicate the position of the peak absolute signal at \(\Delta \tau=0.5\)~ps.
    Note the pump and probe pulses have a temporal width of \(\sigma_t=25\)~fs,
    roughly equivalent to \(\Delta E \sim \hbar/\Delta t  \sim 3\kappa\). For visual clarity
    the plots were individually normalized, with red indicating a positive signal and 
    blue a negative signal.
    }
    \label{fig:sm6}
\end{figure}

\begin{figure}
    \centering
    \includegraphics[width=\linewidth]{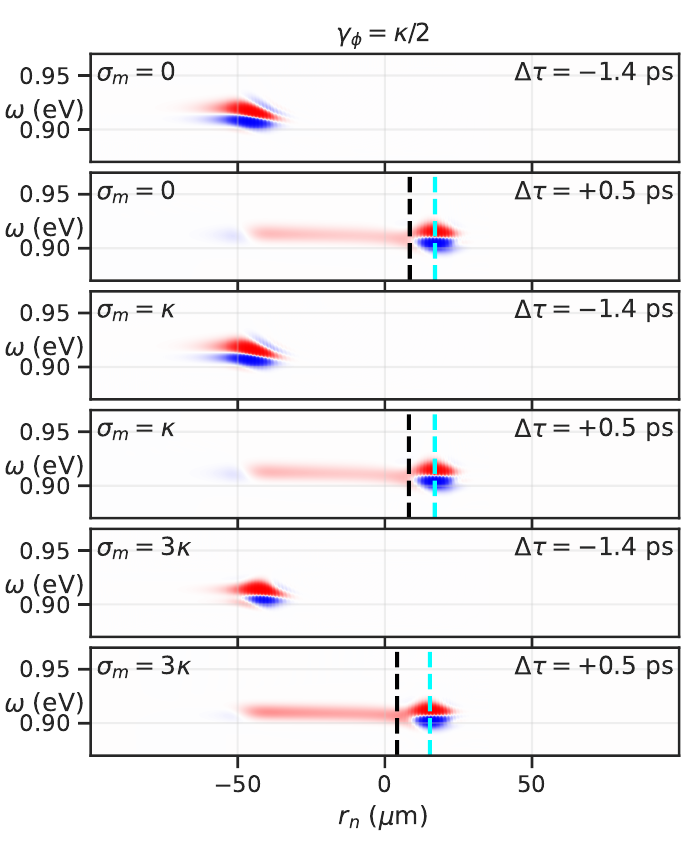}%
    \caption{%
        Differential transmission \(\Delta T_n\) showing position of 
    the LP excitation in the pump-probe experiment at two different times
    with \(k_{p}=-k_{p'}=2 \pi/5\)~\(\mu\text{m}^{-1}\) and \(\gamma_\phi=\kappa/2\),
    for three different disorder widths \(\sigma_m\).
    Dashed cyan lines
    indicate the position of the peak absolute signal at \(\Delta \tau=0.5\)~ps,
    and black dashed lines the root-mean-square position.
    As in \cref{fig:sm6}, the pump and probe pulses have a temporal width of \(\sigma_t=25\)~fs,
    roughly equivalent to \(\Delta E \sim \hbar/\Delta t  \sim 3\kappa\),
    and the plots were individually normalized.
    }
    \label{fig:sm7}
\end{figure}

\Cref{fig:sm6} shows results for the counter-propagating pump-probe 
experiment in the presence of static disorder but without molecular dephasing, \(\gamma_\phi=0\).
This illustrates how, for the model in mean-field theory, energetic disorder in
the molecules alone is not sufficient to couple dark states to the dynamics and 
thereby renormalize the polariton group velocity via the mechanism described in the Letter.
Indeed, moderate disorder  (\(\sigma_m = \kappa \), middle two panels of
\cref{fig:sm6}) has no discernible effect on the differential transmission, whilst
at strong disorder (\(\sigma_m = 3 \kappa \), bottom two panels) there is a change
in spatial pattern of the signal and a small reduction in travel of the signal, attributable
to slight asymmetry of the polariton dispersions at \(k_{p'}=-k_{p}=2 \pi/5\)~\(\mu\text{m}^{-1}\).

Reintroducing molecular dephasing, \cref{fig:sm7}, demonstrates that the transport mechanism 
proposed in the Letter is robust against moderate energetic disorder, with little change 
in the observed signal from \(\sigma_m=0\) to \(\sigma_m=\kappa\). 
At strong disorder (\(\sigma_m=3\kappa\),
bottom two panels of \cref{fig:sm7}),
there does appear to be increased localization of the signal at \(\Delta \tau=-1.4\)~ps and further
a slight reduction in root-mean-square transport velocity (vertical black dashed line).
With the current counter-propagating scheme it is difficult to distinguish whether this is a distinct 
effect from the reduction in group velocities observed in the bottom panel of Fig.~\ref{fig:sm6}.

These results are consistent with low temperature measurements of exciton-polariton transport
in halide perovskite microcavities~\cite{xu2023} indicating that the effects of dynamical disorder
induced from vibronic coupling likely dominate static energetic effects. Nonetheless, it would be interesting
to apply our framework to explore localization effects~\cite{tian2025} arising from static
disorder---and their interplay with exciton hopping---as a route to diffusive transport in future work.
%; nevertheless,
% we note that disorder-induced effects on polariton transport signals have been reported in 
% other theoretical works~\cite{tian2025}. Isolating the effects of disorder---and its interplay 
% with exciton hopping as a route to diffusive transport---is a natural direction for future work.

\end{document}